# Direct Integer Division in RNS and its Hardware Solutions

Author: Eric B. Olsen

Organization: Maitrix, LLC

Email: eric@maitrix.com

## Abstract

Residue Number Systems (RNS) offer efficient modular arithmetic and natural parallelism, but direct integer division in RNS remains a difficult and comparatively underdeveloped operation. This paper builds on the type-II division algorithm of Szabo and Tanaka and reformulates it for more efficient hardware implementation. A principal contribution is the introduction of a power-based RNS, in which moduli are selected as powers of natural primes, increasing dynamic range, improving bit efficiency, and providing greater flexibility for scaling during division. The paper further formalizes three decomposition methods required by the division process: multi-factor scaling for modulus-based division, mixed-radix conversion for base extension and comparison, and a new divisor decomposition method introduced in this work. Each method is supported by mathematical development, including analysis of modulus invalidation during computation. These results simplify the hardware structure of the algorithm and improve its scalability. Supported by hardware diagrams and performance tables, the work advances both the theory and practical implementation of direct RNS division.

## Keywords

## Notation

| Symbol | Description |
|---|---|
| $d_i$ | The $i$-th digit of an RNS number. The index $i$ typically starts from 1. |
| $M_i$ | The $i$-th modulus of an RNS digit. |
| $m_i$ | The $i$-th base modulus of an RNS digit. |
| $N$ | The total number of digits in an RNS word format. |
| $n$ | The fixed-digit width (in bits) to encode in binary a single RNS digit $d_i$. |
| $P$ | The maximum power of a base modulus $m_i$ supported for a given modulus $M_i$. |
| $p$ | The number of powers of base moduli $m_i$ that are factors to a modulus $M_i'$. |
| $x$ | The numerical value represented by an RNS word. |
| $R$ | The entire range of encodings possible for a given RNS format. |
| $R_{\text{int}}$ | The total number of integers that can be encoded by a given RNS word format. |
| $a_i$ | The $i$-th digit in the mixed-radix representation of an RNS number. |
| ‘ | Denotes that a variable has undergone a computation or transformation (e.g., $x'$ is a computed transformation of variable $x$). |

## Introduction

Since the late 1950's, RNS numbers have intrigued researchers due to their carry-free properties. The reason is that RNS numbers enjoy a near constant time of execution versus numeric precision for the integer operations of add, subtract and multiply [1].

However, the usefulness of RNS is impeded by many other arithmetic operations, such as conversion of RNS numbers to binary, comparison of RNS values, digit (base) extension, and overflow detection. In fact, many operations taken for granted when using the binary number system turn out to be quite troublesome when using RNS. Choosing a particular set of RNS moduli with certain properties can solve certain problems, but this approach fails to solve problems in an infinitely extendable number system.

One particularly troublesome RNS operation is arbitrary integer division. In this paper, direct arbitrary RNS division denotes a division method that accepts both the dividend and divisor in RNS format and returns the quotient and remainder in RNS format, without conversion to binary. An additional advantage of the method developed here is that it does not require redundant residue digits, which is important for general-purpose fixed-width RNS hardware where such digits may not exist or be created dynamically. Research on direct arbitrary integer division in RNS is relatively sparse in the broader literature [1], although notable prior works exist [2], [3]. Szabo and Tanaka introduced two such algorithms in [4]. This paper significantly improves one of those algorithms and further discloses a hardware apparatus that performs arbitrary integer division directly in RNS.

The contributions of this paper fall into three categories. First, the paper develops and formalizes mathematical results needed for direct integer division in RNS, including scaling by powers of base moduli, treatment of invalidated moduli, and decomposition methods used by the algorithm. Second, the paper presents a hardware-oriented implementation framework, including RNS digit-processing architectures, control flow, and initial FPGA synthesis results that demonstrate feasibility. Third, the paper identifies several architectural and algorithmic enhancements that are proposed for future investigation and are not claimed as part of the present implementation.

### Benefits of an RNS Divide

Even in binary arithmetic, division is generally the most expensive basic arithmetic operation, requiring substantially greater latency and often more hardware area than addition or multiplication. Despite this, arbitrary division remains indispensable in both general-purpose and specialized computing.

Direct integer division broadens the practical use of RNS in both hardware and software. Many binary algorithms depend on radix-specific shortcuts such as shifting and bit manipulation, which do not translate naturally to RNS. Direct RNS division therefore enables new algorithmic approaches while also supporting accurate fractional representation. More broadly, it expands the role of RNS from specialized arithmetic kernels toward more general computational use.

In addition, a practical direct division capability strengthens the usefulness of RNS as a more complete arithmetic framework. Many important computations require quotient, remainder, scaling, or reciprocal-related operations that are awkward to support if division is unavailable or only available through binary conversion. By keeping these operations within the residue domain, both software algorithms and hardware implementations gain a more direct path toward broader numerical processing while preserving the structural advantages of RNS arithmetic.

This paper presents a practical algorithm for direct integer division in RNS and outlines its hardware implementation. To complement this work, supplemental materials include synthesizable RTL and simulation test benches for arbitrary division on 150-bit RNS numbers, providing an implementation reference for the method presented in this paper.

## Background

Two different types of RNS division algorithms are proposed in [4]. The first "type-I" algorithm uses an iteration involving powers of two. In this algorithm the quotient is obtained by adding a series of RNS values that are pre-computed powers of two. A table of all powers of two for the entire range $R$ of the RNS is typically required but other solutions also exist. A trial quotient is found by starting with the first power of two within a starting range, multiplying the trial quotient with the divisor and determining an error and next range. The process is repeated by adding the next smaller power of two and testing the value against the new range. Several issues are cited with this algorithm. The primary issue is the time-consuming process for computing the large number of RNS comparisons.

The second "type-II" RNS division algorithm is also iterative. It is based on finding an approximation to the divisor which consists of a product of one or more moduli comprising the RNS. Since the approximate divisor is a product of digit moduli, the algorithm can use "scaling" to divide out the dividend using each of its factors. The result of this scaling is indeed a division; it results in an approximate quotient which is then multiplied by the divisor and used to iteratively subtract out a portion of the dividend. The reduced value for the dividend is used in the next iteration using the same approximation for the divisor. The iteration is essentially continued until the dividend goes to zero or is otherwise less than the divisor.

The type-II division algorithm also has several cited issues. One is the need for a look-up table comprising potential modulus combinations to determine a suitable approximate divisor. Moreover, not all sets of moduli satisfy the pre-conditions for selection of the moduli, but many moduli selections will work. A second issue is cited: the indeterminate length of iteration is a function of the values themselves. A third issue is cited here, which is the lengthy number of iterations that may be required if the approximate divisor is not close enough to the actual divisor. In fact, the Type-II algorithm proposed in [4] suffers because it is unclear exactly how the divisor selection is solved for the general case.

A central practical difficulty in direct RNS division is that the required support operations are themselves nontrivial. Comparison, base extension, scaling, and divisor approximation must all be coordinated without relying on conversion to binary or on assumptions that do not hold in a fixed-width general-purpose residue machine. This is one reason direct division has received far less attention than addition or multiplication in RNS. The approach developed in this paper addresses these issues in a unified way by combining power-based moduli, repeated scaling, mixed-radix-based recovery of invalid digits, and a hardware organization designed to support these operations directly within the residue domain.

This paper will describe an improved type-II division algorithm. The improved integer division algorithm will address each of the problems cited above using several novel methods.

## A Quick Review of RNS

A comprehensive review of the residue number system (RNS) is found in [4]. A brief review is included in this paper.

### Notation of Residue Numbers

For this paper, it is helpful to review basic notation for the RNS number system. The RNS number system requires us to select a set of moduli. Therefore, we express a set of $N$ number of moduli $M_i$ within standard parenthesis:

$$\left(M_1,\ M_2,\ M_3,\ \ldots,\ M_N\right) \tag{1}$$

Generally, when a set of moduli is defined, each modulus $M_i$ is chosen to be coprime to all other modulus $M_{j\neq i}$. This ensures that a multiplicative inverse for every $M_i$ with respect to every other modulus $M_{j\neq i}$ exists. This paper assumes only RNS with coprime moduli.

To express an RNS number $x$, we express a set of $N$ number of digits $d_i$ using curly brackets:

$$x=\left\{d_i,\ d_2,\ d_3,\ \ldots,\ d_N\right\} \tag{2}$$

Using this notation, the value of each RNS digit $d_i$ is associated to the digit modulus $M_i$.

### Range of RNS

The total number of unique encodings of an RNS is referred to as the range $R$ of an RNS. In this paper, we only consider positive integer RNS encodings where all moduli $M_i$ are coprime; therefore, the range of positive integers $R_{int}$ of our RNS is given by:

$$R_{int}=R=\prod_{i=1}^{N} M_i \tag{3}$$

### Evaluation of RNS Digits

Of definition of an RNS, to encode a positive integer value $x$ we have:

$$x=\left\{|x|_{M_1},\ |x|_{M_2},\ |x|_{M_3},\ \ldots,\ |x|_{M_N}\right\},\qquad 0\leqslant x<R_{int} \tag{4}$$

Thus, every digit $d_i$ of the RNS value $x$ is obtained by,

$$d_i=|x|_{M_i} \tag{5}$$

### Basic Integer Arithmetic in RNS

An attractive property of RNS is the carry-free feature of integer addition, subtraction and multiplication. In some texts, these arithmetic operations are referred to as parallel array computation (PAC) because each digit can be computed in parallel and without carry [5]. To understand how this works, consider the sum of two RNS numbers, $a$ and $b$:

$$a = \left\{ |a|_{M_1},\ |a|_{M_2},\ |a|_{M_3},\ \ldots,\ |a|_{M_N} \right\} \qquad 0 \le a < R_{int} \tag{6}$$

$$b = \left\{ |b|_{M_1},\ |b|_{M_2},\ |b|_{M_3},\ \ldots,\ |b|_{M_N} \right\} \qquad 0 \le b < R_{int} \tag{7}$$

Then by definition,

$$a+b = \left\{ |a+b|_{M_1},\ |a+b|_{M_2},\ |a+b|_{M_3},\ \ldots,\ |a+b|_{M_N} \right\} \qquad 0 \le (a+b) \le R_{int} \tag{8}$$

Which is equivalent to:

$$a+b = \left\{ \left| |a|_{M_1} + |b|_{M_1} \right|_{M_1},\ \left| |a|_{M_2} + |b|_{M_2} \right|_{M_2},\ \left| |a|_{M_3} + |b|_{M_3} \right|_{M_3},\ \ldots,\ \left| |a|_{M_N} + |b|_{M_N} \right|_{M_N} \right\} \tag{9}$$

Since

$$|a+b|_m = \left| |a|_m + |b|_m \right|_m \qquad \textit{for all } m > 0 \tag{10}$$

Therefore, as seen in equation (9), each RNS digit $d_i$ of the sum $(a+b)$ is computed as the sum of RNS digits $|a|_{M_i}$ and $|b|_{M_i}$ modulo $M_i$. Note the sum $(a+b)$ should be less than the RNS range $R_{int}$ to avoid overflow, which is a difficult scenario to detect in RNS. The same logic used to derive (9) can be applied for integer subtraction and multiplication.

However, for integer division, it is important to note that in general:

$$\left| \frac{a}{b} \right|_{M_i} \ne \frac{|a|_{M_i}}{|b|_{M_i}} \tag{11}$$

Therefore, RNS integer division cannot be performed in a single PAC operation and must be computed using different means. This is the subject of this paper.

## Base Extension of an RNS Digit

The operation of determining the value of a digit $d_i$ for a modulus $m_i$ given the digit value of all other moduli $m_{j\ne i}$ is known as *base extension*. In this definition, "all other" digits $d_{j\ne i}$ comprise a sufficient range to encode the value $x$ to be base extended. In this paper, we refer to "all valid digits" to denote (imply) that a value $x$ is sufficiently encoded. More about the validity of RNS digits and their moduli will be discussed later.

A base extension operation is required for several reasons. One reason is to extend the range of the RNS by adding more digits. In this case, there is a previously encoded value $x$ that should be preserved. After base extension, the added digit is often referred to as redundant, although this is somewhat misleading.

For division, the reason to perform base extension is to recover the value of a digit that has been *invalidated*. Digit invalidation occurs when we divide a residue number by one of its moduli. Division of an RNS number by one or more of its digit moduli is an important operation for division in RNS. In this context, the need to recover a digit $d_i$ of a supported modulus $M_i$ is important so that RNS arithmetic processing can continue. However, the process of base extension is generally considered expensive in terms of processing steps. We will investigate ways to minimize the impact of this important operation later.

### Extending the Range of an RNS Word Format

Given an encoding of a number $x$ using an RNS with $N$ number of co-prime modulus:

$$x = \left\{ |x|_{M_1},\ |x|_{M_2},\ |x|_{M_3},\ \ldots,\ |x|_{M_N} \right\}, \qquad 0 \le x < R_{int} \qquad (12)$$

RNS base extension is the process of adding a new co-prime modulus $M_{N+1}$. The RNS now comprises one more digit. The new digit value follows from the basic definition of an RNS digit value:

$$x = \left\{ |x|_{M_1},\ |x|_{M_2},\ |x|_{M_3},\ \ldots,\ |x|_{M_N},\ |x|_{M_{N+1}} \right\} \qquad (13)$$

After base extension, the previously encoded value $x$ is unchanged even though the previous range $R_{int}$ of the RNS has been extended by a factor of $M_{N+1}$ and is now designated as $R'_{int}$:

$$R'_{int} = \left( \prod_{i=1}^{N} M_i \right) M_{N+1} \qquad (14)$$

Equation 13 makes the process of base extension appear simple; however, in practice the calculation of the digit value $d_{M_{N+1}}$ must be determined from all other digits of the RNS. This is a relatively complex task.

In most texts, the Chinese Remainder Theorem (CRT) is used to find or recover the digit value of the base extended modulus $M_{N+1}$ given the value $x$ in RNS. In this paper, we use mixed-radix conversion as a more direct approach for base extension that is compatible with other required hardware and avoids the need to maintain a redundant digit or find unique solutions based on congruence equations. However, the method of division introduced in this paper is not limited to using mixed-radix conversion to perform base extension. Alternative methods for base extension are an area of study for RNS division algorithms.

### RNS Digit Order Conventions

In this paper, we adopt a left-to-right convention for moduli arranged so the smallest modulus, $M_1$, is positioned on the left. Although RNS does not inherently designate a most significant or "first" digit, an ordering is selected to optimize hardware efficiency. For example, in hardware implementations of mixed-radix conversion, processing RNS digits in order of increasing modulus simplifies the modular subtraction step, making mixed-radix conversion more efficient. Consequently, $M_1$ represents the smallest modulus, $M_p$ denotes the largest, and conversion is performed sequentially from the smallest to the largest modulus.

## RNS Division Overview and Requirements

The integer division introduced in this paper has certain requirements for the selection of the RNS modulus. However, the modulus selection requirements are easily extendable and are efficient for hardware implementation. RNS moduli are chosen using guidelines that maximize the performance benefits of the division algorithm. These guidelines align well with fractional RNS representations developed outside the scope of this paper [6].

### Selection of Modulus

In [7] the ascending sequence of prime numbers starting with two define the modulus of an RNS number system called the *natural residue number system*. We use the natural residue number system to define a set of *base* moduli $m_i$ used in the divide algorithm of this paper; for example, we have the following set of base moduli $m_i$ when $N = 8$:

$$\left( m_1,\ m_2,\ m_3,\ m_4 \dots m_8 \right) = (\ 2,\ 3,\ 5,\ 7,\ \dots,\ 19) \qquad (15)$$

The natural RNS can be infinitely extended; this RNS number system is uniquely defined and provides a framework to derive theoretical relationships for analysis. For example, by selecting *N* residues for our natural RNS, we know exactly which moduli is used. It is intuitively evident that supporting the set of the smallest prime moduli provides a plurality of the most frequently occurring factors for division.

## Power-Based Moduli

The range of the natural RNS can be extended in another way; some of the digit modulus $m_i$ can be raised to some power $P > 1$. This has several benefits. First, it increases the range and bit efficiency of a fixed digit-width RNS machine word, since for hardware design, it is often advantageous to choose a fixed bus-width to store and transfer RNS digits. In terms of the division algorithm, it allows many more scaling factors to be supported. And lastly, it provides a means to delay base extension of the RNS value during division as will be shown later.

An RNS having moduli that are powers of one or more of the natural prime numbers is termed a "power-based" RNS in this paper. For example, the moduli in (15) is extended using powers by:

$$\left( {m_1}^8,\ {m_2}^5,\ {m_3}^3,\ {m_4}^3,\ \dots,\ {m_8}^2 \right) = \left( 2^8,\ 3^5,\ 5^3,\ 7^3,\ \dots,\ 19^2 \right) \qquad (16)$$

To aid the notation of power-based moduli of this paper, the large case $M_i$ is used to denote the maximum supported power of its *base modulus* $m_i$. Since the maximum power $P$ supported of each base modulus $m_i$ may differ, we adopt the following notation and relation:

$$M_i = \prod_{p=1}^{P_i} m_i \qquad (17)$$

During processing with a power-based modulus, it is possible to process using *any power* $p$ of the base modulus $m_i$:

$$\left( m_i^p\ :\ 1 \le p \le P_i \right) \qquad (18)$$

Note that raising a modulus to some power $p$ still maintains the RNS requirement that each modulus is co-prime. Due to the existence of multiplicative inverses described later, this means each power of a specific power-based modulus supports a multiplicative inverse with respect to every other power-based modulus. In the section detailing hardware implementation, a "power valid" count is introduced and used to track a *variable* modulus as one or more base powers are "divided out". These are new and unique ideas for RNS processing central to the division algorithm of this paper.

For a typical hardware implementation, it is desirable to define a fixed-bit width $n$ for encoding all moduli; this will fix the width of data buses and arithmetic multipliers and adders. Thus, an *n*-bit RNS digit encoding dictates the maximum power $P_i$ each base modulus can be extended in (17). By finding the minimum bit width, *n*, to store the largest modulus $M_N$ of the sequence of power-based modulus, the fixed bit width to represent all power-based digits is defined. This will be illustrated in the following paragraphs.

However, we are not limited to the minimum bit width dictated by the power-based modulus. The extent to which the fixed-bit digit width *n* is increased, and the resulting powers of the natural prime sequence extended, is a choice for the hardware designer and indeed an optimization problem of interest.

## Why Support Power-Based Modulus?

Extending the *natural* RNS into *n*-bit power-based digit moduli provides many low value prime factors. For example, the RNS number format example used in this paper will support powers of the base moduli 2, 3, 5, 7, 11, 13, 17, and 19. Increasing the number and power of low-value prime residue digits increases the algorithms' ability to divide the RNS dividend. It will become clear the digit modulus $m=2^n$ is critical to the efficiency of the divide algorithm. In our example, we choose a slightly smaller power of two: $m=2^{(n-1)}$. This design choice will place the base two power modulus into the ascending sequence of all other power-based modulus to help simplify hardware design.

## Hardware Considerations for MOD-9 RNS

To further illustrate, in this paper we use a 9-bit digit width $(n=9)$ and an RNS *machine word* with a total of 18 RNS digits $(N=18)$. This is the design choice used in the MOD-9 hybrid ALU developed by Maitrix, LLC. This choice is made to match the 9-bit multipliers supported by many FPGA devices. For division, we should ensure RNS numbers can be divided out by low value factors; therefore, the choice of power-based modulus uses the first 8 prime numbers as listed in Table 1. As previously mentioned, the order of moduli is not important, however, we choose to order the resulting power-based moduli in ascending order as this can simplify hardware design for mixed-radix conversion.

| $M_1$ | $M_2$ | $M_3$ | $M_4$ | $M_5$ | $M_6$ | $M_7$ | $M_8$ |
|---|---|---|---|---|---|---|---|
| 11^2 | 5^3 | 13^2 | 3^5 | 2^8 | 17^2 | 7^3 | 19^2 |
| 121 | 125 | 169 | 243 | 256 | 289 | 343 | 361 |

*Table 1 – Power-based digit modulus of our RNS register word example*

Not all moduli of our RNS system are required to support powers. Note for example it is not possible to support the next prime digit as a 9-bit power-based modulus since $23^2=529$ and this value cannot be represented using 9-bits ($2^9$ = 512). Therefore, to maximize the efficiency of the RNS, all remaining *non-power-based* digits are prime numbers chosen to exist near the end of the 9-bit digit range (which is $2^9$). Table 2 lists the remaining digit moduli of our example 18-digit RNS word:

| $M_9$ | $M_{10}$ | $M_{11}$ | $M_{12}$ | $M_{13}$ | $M_{14}$ | $M_{15}$ | $M_{16}$ | $M_{17}$ | $M_{18}$ |
|---|---|---|---|---|---|---|---|---|---|
| 457 | 461 | 463 | 467 | 479 | 487 | 491 | 499 | 503 | 509 |

*Table 2 – non-power-based digit moduli of our RNS register word example*

## RNS Bit Representation Efficiency

The representational efficiency of a fixed digit-width RNS, $E_R$, is introduced in [7] and is a measure of the range of the RNS versus the maximum binary range supported by the total number of RNS digit bits. In terms of the notation used in this paper, $n=9$ is the digit bit width, and $N=18$ is the number of RNS digits each supporting a modulus $M_i$. We have:

$$E_R = \frac{\log_2(\prod_{i=1}^{N} M_i)}{n*N} * 100\% \quad (19)$$

For the MOD-9 example modulus used in this paper, the efficiency $E_R$ is:

$$E_R = \frac{151.41}{9*18} * 100\% = 93.46\% \quad (20)$$

Note the MOD-9 supports a binary range of 151.41 bits! Note the effective bit width of the range $R$ of an RNS is never a whole number of binary bits. A representation efficiency of over 93% is reasonable considering the small digit width of 9-bits and the large number of power-based modulus supported.

## RNS Properties

The following sections develop the mathematical properties required for scaling, base extension, and divisor decomposition in the proposed RNS division method.

The binary number system enjoys many advantages because of its *unique properties*. For example, since there exist only two symbols in binary, namely 0 and 1, the product of two binary digits is either 0 or 1. Interestingly, when we multiply a single bit by a much larger binary word, there is no carry from digit to digit! Carry only occurs when we add two binary digits. This and other properties make binary an extremely powerful number system for machine computation.

The residue number system is dramatically different than binary but has its own unique set of properties. The integer division algorithm will take advantage of the following RNS properties. One such property is the ability to easily detect when an RNS number is evenly divisible by one of its moduli.

### Property 1: RNS Divisibility

A non-zero RNS number $x$ is evenly divisible by a modulus $M_i$ having a digit $d_i$ equal to zero. This property is easily verified using the relation for any residue digit, $d_i$, given a non-zero $x$ in RNS:

$$d_i = |x|_{M_i} \quad (21)$$

If $x$ is evenly divisible by $M_i$, we can re-write $x$ as:

$$x = c * M_i \quad (22)$$

and our digit becomes:

$$d_i = |c * M_i|_{M_i} = 0 \quad (23)$$

### Multiplicative Inverse

Some of the most attractive properties of RNS come from RNS being a form of modular arithmetic at the digit level. To maximize efficiency, we make each digit modulus $M_i$ co-prime with respect to all other digit modulus $M_{j \neq i}$. This important restriction allows us to perform inverse multiplication on a digit $d_j$ when we divide by one or more powers of a base modulus $m_i$; to perform this operation, we must know *a-prior* the value of the RNS word $x$ is evenly divisible by one or more powers of the base modulus $m_i$.

## Property 2: Existence of Multiplicative Inverse

Multiplicative inverses are important for RNS division. We will provide an existence definition from [8]:

> *Existence:* There exists an integer denoted $a^{-1}$ *such that* $aa^{-1} \equiv 1 \pmod{n}$ if and only if *a* is co-prime with *n*. This integer is called a *multiplicative inverse* of *a* modulo *n*.

For RNS systems in general, we limit any digit value ***a*** to be less than its modulus $M_i$, such that $0 \le a < M_i$. Therefore, an arithmetic relation with a mod function is used instead of a congruence relation to describe inverse multiplication. The following definition and theorem are reproduced from [4] using a slightly different notation:

*Definition:*

$$\textit{If } 0 \le a < m \textit{ and } |ab|_m = 1,\ a \textit{ is called the multiplicative inverse of } b \bmod m, \textit{ and is denoted by } a = \left|\frac{1}{b}\right|_m.$$

*Theorem:*

$$\textit{The quantity } \left|\frac{1}{b}\right|_m \textit{ exists if and only if } (b, m) = 1 \textit{ and } |b|_m \neq 0. \textit{ In this case } \left|\frac{1}{b}\right|_m \textit{ is unique.}$$

The proof for the theorem is well known and can be found in many texts including [8]. Inverse multiplication using a multiplicative inverse is used to perform scaling and division in RNS. By ensuring the value of an RNS word is evenly divisible by one of its digit modulus, $M_i$, inverse multiplication can achieve a division for all RNS digits $d_{j \neq i}$ by the value $M_i$ since every other digit modulus $M_{j \neq i}$ is coprime, and therefore a multiplicative inverse exists.

To facilitate fast hardware, a multiplicative inverse for each digit modulus $M_i$ with respect to every other modulus $M_{j \neq i}$ is stored in a look-up table (LUT). The hardware is organized so that each digit modulus $M_i$ support a unique LUT containing the multiplicative inverse of each $\left|M_{j \neq i}^{-1}\right|_{M_i}$.

For a power-based modulus, a multiplicative inverse for each modulus power $\left( m_i^k : 1 \le k \le P_i \right)$ with respect to a modulus power $\left( m_{j \neq i}^k : 1 \le k \le P_i \right)$ also exists and is stored in a LUT. Software to generate the required look-up tables can use the extended Euclidean algorithm to determine each multiplicative inverse value. Appendix A includes a table of multiplicative inverses used in this paper and example design.

## Digits for Which Multiplicative Inverse Does Not Exist

In many RNS texts, it is not rigorously discussed what happens to the digit $d_i$ of the RNS value $x$ which is directly undergoing division by its own modulus $M_i$ since for the digit $d_i$ there is no multiplicative inverse with respect to $M_i$.

It was shown in Property 1 that if $d_i = 0$, the RNS value $x > 0$ is evenly divisible by its modulus $M_i$. Because the range of the RNS value $x / M_i$ is decreased by the factor of the modulus value $M_i$, texts often show the digit modulus $M_i$ as simply disappearing. We show in Property 3 mathematics that justify this digit truncation.

In practice, we can choose to consider the digit $d_i$ and its associated modulus $M_i$ as truncated, or we can choose to define the digit $d_i$ and its modulus $M_i$ as undefined or "invalid". In both interpretations, the operation of base extension is used to recover a modulus $M_i$ and its digit $d_i$ value. For hardware implementation, we prefer the latter interpretation, as the hardware supporting the invalid digit modulus still exists. More about this topic will be explored next.

## Relation 1: Modular Arithmetic Division

From the cancellation law for congruences *[9]*,

$$if\ ac \equiv bc(mod\ n) and\ gcd\ (c,n) = 1, then\ a \equiv b(mod\ n) \qquad (24)$$

Which can be used to move modular arithmetic into division:

$$if\ \ b\ divides\ a,\ then\ \left(\frac{a}{b}\right) mod\ n = \frac{(\ a\ mod\ bn)}{b} \qquad (25)$$

## Property 3: Scaling By and Invalidating a Whole Modulus $M_i$

As previously discussed, dividing an RNS value $x$ by a modulus $M_i$ invalidates the digit $d_i$. We can show this using Relation 1 by noting the entire RNS word format is modular with respect to its range $R$:

$$x = |x|_R \qquad 0 \le x < R \qquad (26)$$

If $x$ is evenly divisible by modulus $M_i$, we can re-write $x$ as:

$$x = M_i \cdot c = \left|M_i \cdot c\right|_R \qquad 0 \le x < R \qquad (27)$$

Since $M_i$ evenly divides $R$, using (25) we can write $x$ divided by $M_i$ as:

$$\frac{|x|_R}{M_i} = \left|\frac{M_i \cdot c}{M_i}\right|_{R/M_i} = |c|_{R/M_i} \qquad 0 \le c < \left(\frac{R}{M_i} = R'\right) \qquad (28)$$

This shows that the RNS value $x$ is divided (scaled) by the modulus factor $M_i$ and the modulus of the entire word format of $c$ is scaled by the modulus factor $M_i$. In other words, the resulting value $c$ is the value $x$ directly divided by $M_i$ while the range $R'$ of the resulting RNS format is decreased to $R/M_i$. This can only be true if the resulting RNS number format has its digit $d_i$ and its modulus $M_i$ truncated. We know this since all moduli are co-prime, therefore the reduced range $R'$ of the result of (28) is only achieved if the modulus $M_i$ is eliminated:

$$R' = \frac{R}{M_i} = \left(\prod_{j \ne i} (M_j)\right) \qquad (29)$$

When we divide an RNS number $x$ by a modulus $M_i$, we use inverse multiplication to revert all other digits $d_{j \ne i}$ back to the value they would be before being multiplied by $M_i$. In summary, because the multiplicative inverse for the digit $d_i$ does not exist, we show in (28) that during scaling of $x$ by $M_i$ the digit $d_i$ is effectively truncated, i.e., the modulus $M_i$ is no longer a factor of the modulus $R$ of the RNS word format. In terms of hardware, we prefer to treat the digit and its modulus as "invalid". The digit hardware is ignored and no longer participates in calculations of the (new) RNS word format with range $R^{'}$.

## Property 4: Scaling By and Invalidating One or More Powers of a Base Modulus

Scaling an RNS value by its (whole) modulus $M_i$ is demonstrated in [4]. However, for the division algorithm presented herein we show an RNS value may be scaled by one or more powers of a base modulus $(m_i^p)$. Mathematical reasoning is like Property 3.

It must be known a prior the RNS value $x$ is evenly divisible by one or more powers $p < P_i$ of a base modulus $m_i$; if so, we can write the RNS value $x$ as a product of $p$ powers of a base modulus $m_i$ times a constant $c$. Using Equation 24 we express $x$ to be modular to the range $R$ of the RNS word format:

$$x = m_i^p \cdot c = \left| m_i^p \cdot c \right|_R \qquad 1 \le p < P, \; 0 \le x < R \tag{30}$$

Since $m_i^p$ evenly divides $x$ and $R$, using (25) we can write $x$ divided by $m_i^p$ as:

$$\frac{x}{m_i^p} = \left| \frac{m_i^p \cdot c}{m_i^p} \right|_{R/m_i^p} = |c|_{R/m_i^p}, \qquad 0 \le c < \frac{R}{m_i^p} \tag{31}$$

Equation (31) shows the value x is reduced to the value $c$ (by direct division) and the range $R$ of the original RNS word format is reduced by a factor of $m_i^p$. The result is that the new RNS word format for $c$ has a reduced word modulus or range $R^{'}$ that still contains $P-p$ factors of $m_i$:

$$R^{'} = \left( \prod_{j \ne i} (M_j) \right) \cdot m_i^{P-p} \tag{32}$$

Next, we look at the effect on the digit $d_i$. Because the value of $x$ is evenly divisible by $m_i^p$, we can write the value of $d_i^{'}$ when $x$ is divided by $m_i^p$ as:

$$d_i^{'} = \left| \frac{m_i^p \cdot c}{m_i^p} \right|_{M_i} = \left| \frac{m_i^p \cdot c}{m_i^p} \right|_{m_i^p \cdot m_i^{P-p}} = |c|_{m_i^{P-p}} \tag{33}$$

Equation (33) shows the digit value $d_i$ undergoes direct division by the value $m_i^p$ and its modulus $M_i$ is reduced to $m_i^{P-p}$. We can also show the value of $d_i$ must always be larger than $m_i^p$, so direct division is always possible.

The new number format of the reduced RNS has the original modulus with exception to the modulus $m_i$ of the digit $d_i$ which has been reduced by a factor of $m_i^p$. In (32), all other modulus $M_{j \ne i}$ remain intact. Using definition of multiplicative inverse from Property 2, all other digits $d^{'}_{j \ne i}$ are computed using the multiplicative inverse of $m_i^p$ with respect to the modulus $M_j$ since $m_i^p$ is co-prime with $M_j$:

$$d^{'}_{j \ne i} = \left| \frac{1}{m_i^p} \right|_{M_j} \cdot \left| m_i^p \cdot c \right|_{M_j} = |c|_{M_j} \tag{34}$$

In the version of the divide algorithm presented, to enhance speed of division (in most cases) the algorithm will delay the base extend operation to recover invalidated moduli until all powers of the base modulus $m = 2$ are invalidated, i.e., the entire modulus $M = 2^P$ is invalid. However, there are variations to the algorithm where base extension is performed when any modulus of a set of the lowest value base moduli is invalidated. This is an important area of study for optimization of the RNS divide algorithm.

## RNS Decompositions

This paper refers to the conversion of an RNS value to mixed-radix format using the general term "decomposition". We say the RNS value is decomposed into its mixed-radix form. In this way, our terminology is with respect to RNS, not mixed-radix or binary number format.

Decomposition is a fundamental concept in RNS processing, as it is in any number system. For example, a fixed-radix number can be expressed as a summation, or base expansion [10], of weighted digits. Binary arithmetic frequently takes advantage of this *decomposition* property. Unlike fixed-radix systems, however, RNS is not a weighted or positional number system, which makes its decomposition algorithms distinct. Despite these differences, RNS decompositions are crucial for various computational processes and warrant detailed study.

Three types of RNS decomposition are used by the division algorithm described herein.

1. **Multi-factor scaling**: The division of an RNS number by one or more base modulus.
2. **Mixed-radix conversion**: Converting an RNS number to a mixed-radix representation.
3. **Divisor decomposition**: A new class of RNS decomposition techniques defined herein.

In summary, these three types of RNS decomposition: multi-factor scaling, mixed-radix conversion and divisor decomposition are crucial to the RNS integer division algorithm.

### Multi-Factor Scaling

As demonstrated in [4], basic scaling of an RNS value $x$ is a fundamental low-level operation in many RNS arithmetic algorithms. For the division method of this paper which supports power-based modulus, we extend scaling to include scaling by one or more powers of a base modulus. In addition, we show the obvious extension of performing more than one scaling operation on an RNS value provided the RNS value is evenly divisible by each scaling factor.

It has been shown in property 4 that scaling an RNS value $x$ by $m_i$ represents a full division of the value $x$ by $m_i$. Furthermore, scaling requires the RNS value $x$ to be evenly divisible by the scaling factor $m_i$. Applying a plurality of MOD (%) functions helps us determine if the RNS value $x$ is evenly divisible by one or more powers of $m_i$ as a pre-requisite to scaling. A side effect of scaling is the digit modulus $M_i$ is also scaled or invalidated. Scaling also relies on the existence of a multiplicative inverse for each modulus $m_{j \neq i}$ with respect to $m_i$ as explained in Property 2; multiplicative inverses are typically stored in a LUT for the division algorithm of this paper. MOD functions are typically implemented using LUTs since the MOD function is with respect to a constant modulus value.

## Example of RNS Multi-Factor Scaling

Figure 1 illustrates the sequential scaling of an RNS value $x = 6000$ by the factors 125, 3, and $2^4$. Using the power-based modulus of Table 1, the example of Figure 1 first shows scaling by a full digit modulus $M_2=125$, then shows scaling by the base modulus $m_4$=3, and finally scales by four powers of the base modulus $m_5=2^4$. The multiplicative inverse for each RNS digit $d_i$ of $x$ for each scaling operation is also shown in Figure 1. These multiplicative inverses may also be found in the table of Appendix A.

In Figure 1 step 0 shows a starting value of x=6000. Note the digit $d_2$=0 since the value 6000 is evenly divisible by the modulus $M_2$=125. This fact is easily identified by inspection alone. The multiplicative inverses of $M_2$=125 with respect to each other digit modulus $M_i$ is listed in step 1. The result of multiplying each multiplicative inverse by the respective RNS digit $d_i$ mod $M_i$ is shown in Step 2. Note the digit $d_2$ is now marked with an asterisk signifying it is undefined since the modulus $M_2$ has been invalidated.

The digit $d_2$ and its modulus $M_2$ can be thought of as being "divided out". In this paper, we show an asterisk to denote the digit $d_2$ is "invalid" or undefined. The term invalid is useful for hardware implementations since something will exist in the digit register but must be ignored. To recover the value of this register, a base extension operation must be performed, but it is not necessary to perform base extension to continue scaling by another base or full modulus.

In step 2, it is "detected" that the base modulus $m_4$=3 evenly divides the new value of $x$=48. In a later section covering hardware, we will show how this case is performed by LUT's and logic which checks if a mod function by one or more powers of the base modulus $m_i$ is zero. Since the digit $d_4$ is evenly divisible by its base modulus $m_4$, it will be directly divided. This division by a constant is performed using LUTs. Step 3 shows the multiplicative inverse of $m_4$ with respect to all other modulus $M_i$ in preparation for inverse multiplication in Step 4.

Once again, the result of multiplying each multiplicative inverse by its respective RNS digit $d_i$ mod $M_i$ is shown in step 4. Also shown in Step 4 is the reduction of the range of the modulus $M_4$ by the number of the powers being "divided out". In this case, only one power of $m_4$ is divided out by inverse multiplication; therefore, the total modulus $M_4=3^5=243$ is reduced by one power to $M_4=3^4=81$. The theory behind this process was explained in Property 3.

In Step 5, it is also detected that the base modulus $m_5$=2 divides the digit $d_5$=16 evenly. Moreover, up to 4 powers of $m_5$ divide the digit $d_5$ evenly. To enhance efficiency, it is more desirable to divide the digit $d_5$ by four powers instead of one power of $m_5$. In general, the maximum number of powers that can be divided out depend on the maximum number of powers that evenly divide the digit $d_i$ and is further limited by the maximum number of *available* powers of the modulus $M_i$; we cannot divide out more powers than the modulus supports in a single scaling operation. Note in the general case, the modulus $M_i$ may have undergone a reduction in power due to a previous operation.

In step 5 the values of the multiplicative inverse of $m_5^4$ with respect to all other digit moduli $M_i$ are shown. The result of the modular multiplication of each multiplicative inverse to their respective digit $d_i$ is shown in Step 6. The result shows the value has been reduced to $x = 1$ in this example. Also shown in blue is the reduction of the modulus $M_5$ by four powers of the base modulus $m_5$.

In summary, multi-factor scaling is an important operation for reducing values in RNS. Multi-factor scaling is used throughout the division algorithm of this paper. As the example intuitively shows, supporting many moduli $M_i$ which are the product of small primes ($m_i$) provides on average more opportunities to scale the value of an arbitrary RNS value.

| | Digit modulus | $M_1$ = 121 ($11^2$) | $M_2$ = 125 ($5^3$) | $M_3$ = 169 ($13^2$) | $M_4$ = 243 ($3^5$) | $M_5$ = 256 ($2^8$) | $M_6$ = 289 ($17^2$) | $M_7$ = 343 ($7^3$) | $M_8$ = 361 ($19^2$) | |
|---|---|---|---|---|---|---|---|---|---|---|
| Step | Digit index (*i*) | $d_1$ | $d_2$ | $d_3$ | $d_4$ | $d_5$ | $d_6$ | $d_7$ | $d_8$ | Notes |
| 0 | Starting value | 71 | 0 | 85 | 168 | 112 | 220 | 169 | 224 | $x=6000_{10}$ |
| 1 | Multiplicative inverses for $M_2$ | 91 | * | 96 | 35 | 213 | 37 | 118 | 26 | x is divisible by $M_2$=125 |
| 2 | Multiply by $\left\|\frac{1}{M_2}\right\|_{M_i}$ | 48 | * | 48 | 48 | 48 | 48 | 48 | 48 | Result of scaling x by 125 |
| 3 | Multiplicative inverses for $m_4$ | 81 | * | 113 | * | 171 | 193 | 229 | 241 | x is divisble by $m_4$=3 |
| 4 | Multiply by $\left\|\frac{1}{m_4}\right\|_{M_i}$ | 16 | * | 16 | 16 | 16 | 16 | 16 | 16 | Result of scaling x/125 by 3 |
| | Update modulus $M_4$ | | | | $M_4$ = 81 | | | | | |
| 5 | Multiplicative inverses for ${m_5}^4$ | 53 | * | 74 | 76 | * | 271 | 193 | 158 | X is divisible by $(m_5)^4$=16 |
| 6 | Multiply by $\left\|\frac{1}{{m_5}^4}\right\|_{M_i}$ | 1 | * | 1 | 1 | 1 | 1 | 1 | 1 | Result of scaling x/(125*3) by $2^4$ |
| | Update modulus $M_5$ | | | | | $M_5$ = 16 | | | | |

*Figure 1 - Example of multi-factor scaling in RNS*

## Mixed-Radix Conversion

Mixed-radix conversion is used to convert an RNS number into an equivalent mixed-radix number (MRN) [4]. Like all number systems, the MRN system possesses its own unique properties. Fundamentally, mixed-radix numbers are weighted and positional, therefore comparison is relatively straight-forward. It follows that each mixed-radix digit represents a value, such that all mixed-radix digits of a mixed-radix number may be viewed as a summation of weighted products. For an RNS value $x$ having the moduli $\{m_1, m_2, m_3, \ldots, m_N\}$, and assuming a mixed-radix conversion in the order of ascending moduli index as shown, the MRN has a form given by:

$$x = a_1 + a_2 m_1 + a_3 m_1 m_2 + a_4 m_1 m_2 m_3 + \ldots, + a_p m_1 m_2 m_3, \ldots, m_{N-1} \quad (35)$$

In (35), the value $a_1$ is the least significant mixed-radix digit, and $a_p$ is the most significant digit. Like a fixed-radix number, a MRN is a summation of weighted mixed-radix digits where each weight is the product of all preceding digit radix $m_i$. Each mixed-radix digit has a valid range equal to the range (modulus) of the associated RNS digit of the same index.  More about mixed-radix conversion is included in [4].  Interestingly, mixed-radix numbers do not multiply or divide easily, and very little research in this area is available.

When mixed-radix conversion is performed, the digits are generated starting with the least significant digit.  The convention used in this paper is that mixed-radix digits are ordered from least significant to most significant in a non-traditional "reverse" order of significance, i.e., from left to right.  This reflects the order of mixed-radix digit generation in a left to right format.  Comparison in hardware must support a comparison starting with the least significant mixed-radix digits first.

In the RNS hardware described in this paper, mixed-radix conversion is employed for two primary purposes: (1) comparison of RNS values and (2) base extension of RNS values. Additionally, the principles of mixed-radix conversion provide insight into the sequence of RNS scaling operations performed in the division algorithm. However, during RNS hardware processing, it is unnecessary to fully store a mixed-radix number. Instead, each mixed-radix digit is generated on-the-fly, used immediately to process the RNS value, and then discarded.

In many papers, the Chinese Remainder Theorem (CRT) is used to perform base extension, however, the mixed-radix conversion is more practical since it shares common hardware with the scaling and divisor decompositions to be introduced later.  Another major disadvantage of CRT for RNS processing is we cannot guarantee the RNS number undergoing division includes a redundant digit.  This requirement cannot be met in general if we expect general-purpose RNS computation.  Moreover, the mixed-radix conversion can adapt to any RNS number format, since one or more moduli may be invalidated.

The sequential processing of mixed-radix conversion is often cited as a disadvantage; however, we are performing a multi-cycle division in this paper.  In this context mixed-radix conversion adds execution cycles for the integer divide.  However, this paper will show how the frequency of mixed-radix conversions is decreased, i.e., by 1) delaying the need to base extend a value using power-based moduli, and 2) by base extending multiple digits simultaneously in the same operation.  Furthermore, as the division proceeds, the values being operated on decrease in magnitude, and this shortens the number of cycles for mixed-radix conversion.

## Mixed-Radix Conversion Algorithm

In mixed-radix decomposition, an RNS value is iteratively reduced by subtracting a digit $d_i$ and then dividing (scaling) the result by the corresponding digit modulus $M_i$. This process repeats for each digit until the remaining RNS value reaches zero. The division step, i.e., dividing by a modulus (or a power of a base modulus), is effectively performed by multiplying each RNS digit by an appropriate multiplicative inverse. In each case, the operation is a modular multiplication modulo $M_i$. The range *R* of the RNS number system containing $x$ decreases by a factor $M_i$.  To ensure valid modular division, the RNS value $x$ must first be adjusted to be divisible by the chosen digit modulus, which is achieved by subtracting the digit $d_i$ as shown:

$$x' = \left|\frac{x-d_i}{M_i}\right|_{R/M_i}$$

Mixed-radix conversion can be applied in any order of RNS digits, each order generating a distinct mixed-radix number system. When the dividend and divisor are converted using the same digit order, the resulting mixed-radix digits can be directly compared thereby performing the operation of comparison in RNS. For tasks such as base extension, any RNS digit order can be used for the mixed-radix conversion. Figure 2 is an example of a mixed-radix conversion.

## *Mixed-Radix Conversion Example*

In step 0 of Figure 2 the starting value of x=123,456 is shown in RNS format; the format of the RNS number system is an eight-digit number system with the moduli listed in the top header. In step 1, the RNS number *x = {36, 81, 86, 12, 64, 53, 319, 355}* is subtracted by the value of the first digit $d_1$=36. The choice of digit to start with is arbitrary, but for hardware implementation, an order is chosen; in this case, we are processing $d_1$ through $d_8$ in ascending order.

| | Digit modulus | $M_1$ = 121 | $M_2$ = 125 | $M_3$ = 169 | $M_4$ = 243 | $M_5$ = 256 | $M_6$ = 289 | $M_7$ = 343 | $M_8$ = 361 | |
|---|---|---|---|---|---|---|---|---|---|---|
| Step | Digit index | $d_1$ | $d_2$ | $d_3$ | $d_4$ | $d_5$ | $d_6$ | $d_7$ | $d_8$ | Notes |
| 0 | Starting value | 36 | 81 | 86 | 12 | 64 | 53 | 319 | 355 | $x=123456_{10}$ |
| 1 | Subtract by $d_1$=36 | 0 | 45 | 50 | 219 | 28 | 17 | 283 | 319 | Mixed radix digit $a_1$ = 36 |
| 2 | Multiply by $\left\|\frac{1}{M_1}\right\|_j$ | * | 20 | 6 | 48 | 252 | 153 | 334 | 298 | Dividing by 121; $d_1$ is invalidated |
| 3 | Subtract by $d_2$=20 | * | 0 | 155 | 28 | 232 | 133 | 314 | 278 | Mixed radix digit $a_2$ = 20 |
| 4 | Multiply by $\left\|\frac{1}{M_2}\right\|_j$ | * | * | 8 | 8 | 8 | 8 | 8 | 8 | Dividing by 125; $d_2$ is invalidated |
| 5 | Subtract by $d_3$=8 | * | * | 0 | 0 | 0 | 0 | 0 | 0 | Mixed radix digit $a_3$ = 8 |
| 6 | Zero detected | * | * | 0 | 0 | 0 | 0 | 0 | 0 | STOP |

*Figure 2 - Mixed-radix Algorithm Example*

In step 2, each digit result of step 1 is then multiplied by the multiplicative inverse with respect to $M_j$. Refer to the table in Appendix A for the value of each multiplicative inverse value as they differ for each modulus $M_j$. Note there is no multiplicative inverse that exists for digit $d_1$=0, as described by Property 2; therefore, the digit position $d_1$ is invalidated as shown by an asterisk.

Steps 1 and 2 are repeated for each digit $d_i$ until step 5 where we note the subtraction results in zero for every *valid* (remaining) digit. The algorithm terminates at step 6 when the all-zero state is detected. The result of each subtraction yields a mixed-radix digit. The mixed-radix digits $a_i$ of Figure 2 are weighted, therefore using Equation 1 the mixed-radix number may be expanded and recalculated as a decimal value:

$$x = 36 + 20(121) + 8(121 \cdot 125) = 123,456 \qquad (36)$$

## Mixed-Radix Conversion for Base Extension

Base extension is the process of recovering RNS digits (and their respective modulus) that are invalidated by scaling operations. In the division algorithm of this paper, mixed-radix conversion is used to perform base extension.

The base extension algorithm presented can extend any number of invalid RNS digits simultaneously and “recomposes” each invalid digit during the mixed-radix conversion. To support this, the mixed-radix conversion can start at any digit position but will ignore invalid (skipped) digits. During mixed-radix conversion, mixed-radix digits are generated and then multiplied with that digit’s weight. A unique feature of the base extension procedure is each invalid digit is computed as a running sequence of modular multiplications. See (35) for the expression of the value of a mixed-radix number.

### Base Extension Using Mixed-Radix Conversion Example

Figure 3 is included to illustrate the base extension algorithm used by the division algorithm. This example uses the same starting value x=123,456 as in Figure 2; however, the number of digits tabulated is reduced for brevity. (Note the “range” of digit moduli processed in Figure 3 is still enough to cover the range of the value of x, since $125 \cdot 169 \cdot 243 = 5,133,375$).

In Step 0, the RNS starts with an invalid digit $d_1$ that will be recovered. The column label *Base extend* shows the mixed-radix digit recombination calculation of (36). The column labeled $|weight|_{121}$ lists the running mixed-radix weight calculation, and the column labeled $|d_1|_{121}$ lists the running calculation for the recovered digit $d_1$. Note these last columns perform arithmetic in a modular fashion, using the modulus $M_1$=121.

In Figure 3, the mixed-radix conversion starts with digit $d_2$ since $d_1$ is invalid. The mixed-radix conversion proceeds normally from left to right. In general, but not shown, if an invalid digit is encountered, it is skipped. The mixed-radix number system weight is processed in a sequential manner and in a modular fashion as the conversion proceeds; therefore, the weight is always bounded by the modulus $M_1$=121. In a similar manner, the digit $d_1$ is sequentially processed and always remains bounded by $M_1$. When the resulting RNS value goes to zero in Step 5, the last modular computation is performed to provide the recovered value $d_1 = 36$. Note the number of operands for the computation of $d_1$ in each step will never exceed three. This technique is synthesized into hardware in a later section.

| | Digit modulus | $M_1$ = 121 | $M_2$ = 125 | $M_3$ = 169 | $M_4$ = 243 | | | | |
|---|---|---|---|---|---|---|---|---|---|
| Step | Digit index | $d_1$ | $d_2$ | $d_3$ | $d_4$ | Notes | Base extend | $\|weight\|_{121}$ | $\|d_1\|_{121}$ |
| 0 | Starting value | * | 81 | 86 | 12 | $x=123456_{10}$ | | | |
| 1 | Subtract by $d_2$=81 | * | 0 | 5 | 174 | Mixed radix digit $a_1$ = 81 | hold 81 | 1 | |
| 2 | Multiply by $\left\|\frac{1}{M_2}\right\|_j$ | * | * | 142 | 15 | Dividing by 125; $d_2$ is invalidated | add (1*81) | | \|81\| |
| 3 | Subtract by $d_3$=142 | * | * | 0 | 116 | Mixed radix digit $a_2$ = 142 | Hold 142 | \|125\|=4 | |
| 4 | Multiply by $\left\|\frac{1}{M_3}\right\|_j$ | * | * | * | 5 | Dividing by 169; $d_3$ is invalidated | Add (142*125) | | \|81+4*142\| = 44 |
| 5 | Subtract by $d_4$=5 | * | * | * | 0 | Mixed radix digit $a_3$ = 5 | Hold 5 | \|4*169\|=71 | |
| 6 | Zero detected | * | * | * | 0 | STOP | Add (5*125*169) | | \|44+5*71\| = 36 |

*Figure 3 – Mixed-radix conversion used to recover the value of $d_1$*

To better understand the base extension algorithm used, the value of the recovered digit $d_1$ can be written as a sequence of running modular additions and multiplications:

$$d_1 = \left| \left| |81|_{121} + \left| 142 \cdot |125|_{121} \right|_{121} \right|_{121} + \left| 5 \cdot |125 \cdot 169|_{121} \right|_{121} \right|_{121} \quad (37)$$

In summary, the base extension algorithm will recover any number of invalid RNS digits simultaneously. For example, any invalid digit can be base extended by replacing the mod function in (37) with that digits modulus. The more invalid digits to be recovered, the less number of steps in the mixed-radix conversion because there are less valid digits to compute. This is also true as the value to be base extended is smaller in magnitude. This reveals why delaying base extension is advantageous to the divide algorithm. We will see later that power-based moduli play a key role in the delay of base extension.

## Divisor Decomposition

The divisor decomposition algorithm is unique to this paper. This decomposition is comprised of a series of scaling and incrementing operations until the divisor equals one. The divisor decomposition algorithm as described herein requires the existence of a base modulus =2 to be supported in the RNS number format. Several variations to the divisor decomposition algorithm exist, however, we will introduce a basic version using an "add one" strategy.

The decomposition of an RNS divisor $Y$ is described by a series of values $Y_i$, so that:

$$Y_{i+1} = \begin{cases} \dfrac{Y_i}{D_i}, & if\ \exists D_i\ such\ that\ D_i \mid Y_i \\ Y_i + 1, & otherwise \end{cases} \quad (38)$$

and

$$Y_0 = Y \quad (39)$$

and

$$D_i = \prod_{j\,:\,d_j\ mod\ m_j^k = 0} \left\{ m_j^k : k \geq 1 \right\} \quad (40)$$

During the decomposition, the starting value $Y = Y_0$ is tested for all digits $d_j$ that are zero or evenly divided by one or more powers of its base modulus $m_j$. The value $Y_0$ is divided by the product of all such modulus $m_j$, however, in practice the value $Y_1$ is obtained after successive divisions by each modulus $m_j$ using the scaling algorithm. It is also possible to support in hardware a divide by some power of the base modulus, $m_j^k$ in a single step provided a larger inverse LUT is supported. If no modulus which evenly divides $Y_i$ is found, then $Y_{i+1}$ is obtained by incrementing $Y_i$ by one. In this case it is known that $Y_{i+1}$ must have its base modulus $m = 2$ evenly divisible by one or more powers of 2. The iterative decomposition of $Y$ continues until the value $Y_{i+1} = 1$.

During the decomposition of $Y$, it may become necessary to base extend $Y_i$ to maintain the valid digits of the moduli that undergo scaling. For example, it is important to base extend if the base=2 modulus becomes invalid, otherwise, it is impossible to guarantee at least one modulus that divides $Y_i$ evenly after incrementing $Y_{i-1}$. In this basic decomposition algorithm, it is generally important that only one increment be allowed before $Y_{i-1}$ is divided using the scaling algorithm; however, this is not a strict requirement in every case and is therefore an area for more research. Note in (38), there is no rule specified as to when and how often to base extend $Y_i$, and therefore the decomposition sequence is not unique for a given set of divisor moduli until this rule is defined. The rules we employ to base extend any $Y_i$ can vary and this affects the efficiency of the divide algorithm as will be explained later.

The goal of the divisor decomposition algorithm is to obtain a value $\widehat{Y}$, where:

$$\widehat{Y} = \prod_i D_i \tag{41}$$

The value $\widehat{Y}$ is a product of actual and/or 'approximate' factors of $Y$. All factors of $\widehat{Y}$ are from the set of base moduli and their powers. Because of the increment step in the iteration of (38),

The value $\widehat{Y}$ is either equal to or greater than the value $Y$. The goal is that $\widehat{Y}$ closely approximates $Y$.

$$\widehat{Y} \geq Y \tag{42}$$

## Divisor Decomposition Example

Figure 4 illustrates an example divisor decomposition of the value *x=123456* starting in Step 0. It is detected that the value 123456 is evenly divisible by $m_4=3$ in Step 2 which lists the multiplicative inverses of 3 with respect to each digit modulus. The result of scaling the value x=123456 by 3 is shown. Also in Step 1, the result of dividing the value 123456 by 3 is that the modulus $M_4$ is reduced by one power of 3 to a value of 81.

In Step 2, the resulting value of x'=41152 is detected to be divisible by 64. The multiplicative inverse of 64 with respect to each digit modulus is listed. The result of the modular multiplication with each inverse yields the RNS value 643 as shown in the next row. Also in Step 2, modulus $M_5$ is reduced by 6 powers of 2 ($2^6$=64) to a value of 4 (i.e., $2^{8-6}$). In Step 3, it is detected that the value 643 is not divisible by any base modulus. Therefore, the value 643 is incremented by one to 624 which is divisible by 4. The increment operation is described in (38).

In Step 4, the value 624 is divided by 4. Note that the digit modulus $M_5$ is now depleted; all powers of the original modulus $M_5$ are invalid. Because the entire modulus $M_5$ is invalid, an asterisk is used to denote the digit $d_5$ is invalid. Even though the RNS value now has an invalid base modulus of two, the decomposed value is evenly divisible by the base modulus $m_7$=7. Therefore, in Step 5 the value 161 is divided by 7 to become 23. The modulus $M_7$ is reduced to a value of $7^2$.

In Step 6, the decomposed value x''' is not divisible by any base modulus and furthermore the base two modulus $M_5$ is undefined. In this case, a base extension operation is required to recover the value of the $M_5$ modulus. This is necessary since we must increment the value of x'''=23 to 24 in Step 7 to make it evenly divisible by the base modulus $m_4$=3 and $m_5=2^3$. In Step 8, value 24 is divided by the value of 3 resulting in a value of 8. The modulus $M_4$ is reduced by one power of three to $3^4$. And finally, in Step 9 the value of 8 is divided by three powers of two to bring the decomposed value of x to equal one. The decomposition stops at this point. The total number of factors found by the decomposition is 3 x 64 x 4 x 7 x 3 x 8 = 129024.

| | Digit modulus | $M_1$ = 121 ($11^2$) | $M_2$ = 125 ($5^3$) | $M_3$ = 169 ($13^2$) | $M_4$ = 243 ($3^5$) | $M_5$ = 256 ($2^8$) | $M_6$ = 289 ($17^2$) | $M_7$ = 343 ($7^3$) | $M_8$ = 361 ($19^2$) | |
|---|---|---|---|---|---|---|---|---|---|---|
| Step | Digit index (*i*) | $d_1$ | $d_2$ | $d_3$ | $d_4$ | $d_5$ | $d_6$ | $d_7$ | $d_8$ | Notes |
| 0 | Starting value | 36 | 81 | 86 | 12 | 64 | 53 | 319 | 355 | $x=123456_{10}$ |
| 1 | Multiplicative inverses for 3 | 81 | 42 | 113 | * | 171 | 193 | 229 | 241 | x is divisible by $m_4$=3 |
| | Multiply by $\left\|\frac{1}{3}\right\|_{M_i}$ | 12 | 27 | 85 | 4 | 192 | 114 | 335 | 359 | Result of scaling x by 3. 123456/3= 41152 |
| | Update modulus $M_4$ | | | | $M_4$=81 | | | | | |
| 2 | Multiplicative inverses of 64 | 104 | 84 | 103 | 19 | * | 140 | 134 | 220 | x' is divisble by $m_5^6$= 64 |
| | Multiply by $\left\|\frac{1}{64}\right\|_{M_i}$ | 38 | 18 | 136 | 76 | 3 | 65 | 300 | 282 | Result of scaling x' by 64. 41152/64= 643 |
| | Update modulus $M_5$ | | | | | $M_5$ = 4 | | | | |
| 3 | Increment x' | 39 | 19 | 137 | 77 | 4 | 66 | 301 | 283 | x'' is now divisible by 4. |
| 4 | Multiplicative inverses of 4 | 91 | 94 | 127 | 61 | * | 217 | 86 | 271 | Result of scaling x'' by $2^2$ 644/4=161 |
| | Multiply by $\left\|\frac{1}{4}\right\|_{M_i}$ | 40 | 36 | 161 | 80 | * | 161 | 161 | 161 | |
| 5 | Multiplicative inverses of 7 | 52 | 18 | 145 | 58 | * | 124 | * | 258 | x''' is divisible by 7. |
| | Multiply by $\left\|\frac{1}{7}\right\|_{M_i}$ | 23 | 23 | 23 | 23 | * | 23 | 23 | 23 | Result of scaling x''' by 7 161/7=23 |
| | Update modulus $M_7$ | | | | | | | $M_7 = 7^2$ | | |
| 6 | Base extend x''' | 23 | 23 | 23 | 23 | 23 | 23 | 23 | 23 | Base extend x''' recovers $M_5$ |
| 7 | Increment x''' | 24 | 24 | 24 | 24 | 24 | 24 | 24 | 24 | x'''' now divisible by 3 & 8 |
| 8 | Multiplicative inverses of 3 | 81 | 42 | 113 | * | 171 | 193 | 229 | 241 | Result of scaling x'''' by 3 24/3=8 |
| | Multiply by $\left\|\frac{1}{3}\right\|_{M_i}$ | 8 | 8 | 8 | 8 | 8 | 8 | 8 | 8 | |
| | Update modulus $M_4$ | | | | $M_4 = 3^4$ | | | | | |
| 9 | Multiplicative inverses of 8 | 106 | 47 | 148 | 71 | * | 253 | 43 | 316 | Result of scaling x''''' by 8 8/8=1 |
| | Multiply by $\left\|\frac{1}{8}\right\|_{M_i}$ | 1 | 1 | 1 | 1 | 1 | 1 | 1 | 1 | |
| | Update modulus $M_5$ | | | | | $M_5 = 2^5$ | | | | |

*Figure 4 - Example divisor decomposition*

## General Division Algorithm

The integer division algorithm is based on a recurrence equation introduced in [4] but with several distinct differences. Unlike [4] the recurrence equation includes a unique approximation to the divisor derived using the divisor decomposition algorithm introduced earlier.

The primary objective is to compute the integer division:

$$Z = \left\lfloor \frac{X}{Y} \right\rfloor \tag{43}$$

We first obtain a value $\widehat{Y}$ using the divisor decomposition algorithm such that:

$$\widehat{Y} \geq Y \tag{44}$$

Then the recurrence equation is given by:

$$X_i = X_{i-1} - YZ_i, \quad X_0 = X \tag{45}$$

and

$$Z_i = \left\lfloor \frac{X_{i-1}}{\widehat{Y}} \right\rfloor \tag{46}$$

to obtain $Z_1, Z_2, \ldots Z_n$

The iteration procedure is continued until $Z_n = 0$.

The result of the integer division is then given by:

$$Z = \left\lfloor \frac{X}{Y} \right\rfloor = \sum_{i=1}^{n-1} Z_i + Z_a \tag{47}$$

where:

$$Z_a = \begin{cases} 1, \ \textit{if} \ X_n \geq X \\ 0, \ \textit{otherwise} \end{cases} \tag{48}$$

The general division recurrence of (45) and (46) is combined with RNS decomposition methods explained earlier, such as the divisor decomposition, and base extension using mixed-radix conversion. The divisor decomposition is crucial; not only does it serve to represent the divisor $\hat{Y}$, but the divisor decomposition of $Y$ occurs by a series of divisions (scaling) by one or more powers of the base moduli as expressed in (38). During each of these decomposition divisions, the same division also occurs for the dividend $X$ to compute a term of the recurrence of (46).

Base extension also plays a key role. During the process of repeated scaling by one or more powers of a base modulus, it is possible the base modulus $m = 2$ is invalidated; moreover, there may be no other opportunities to further scale the divisor because no other divisor base moduli are "zero", i.e., no other base modulus evenly divides the divisor. When this occurs, the division algorithm will base-extend all digits that are invalid. All invalid digits are base extended in the same mixed-radix conversion resulting in a *normalized RNS word* format, i.e., all digit moduli are returned to their full power, and their digits are restored.

After base-extension, the divisor is again ready to be divided by new base moduli that are "zero", or if no zeros exist, the divisor is incremented according to (38). Therefore, base-extension of invalid RNS digits plays a critical yet hidden role in the recurrence equations presented. (Note: the term "zeroes" is used in this paper to describe if an RNS value is evenly divisible by one or more powers of at least one of its base modulus).

## Divide Algorithm Control State Diagram

The state diagram flow of Figure 5 illustrates an abbreviated state transition as implemented in the code materials and represents the implementation of the integer division control module of Figure 14. Refer to the code materials in Appendix B for the detailed state transition implemented. The materials are preliminary as many opportunities exist to enhance the division algorithm.

The state transition diagram starts with an IDLE state which waits for the start signal to be asserted. Upon the start signal assertion, the finite state machine (FSM) transitions to the **LOAD_INIT** state. In the LOAD_INIT state, the DENOM RNS register is loaded with the divisor, and NUMER RNS register is loaded with the dividend. After loading the RNS registers, three conditions are checked. 1) If the divisor is zero, a divide by zero error condition is entered. 2) If the divisor (DENOM) is divisible by one or more powers of a base modulus, then the DIV_DENOM state is entered. 3) if the divisor is not divisible by any of its power-based modulus, the divisor is incremented by one in state **INC_DENOM**. This ensures DENOM is evenly divisible by at least the m=2 modulus. The FSM will then transition to the **DIV_DENOM** state.

The **DIV_DENOM** state controls the process of repeatedly dividing the divisor by any base modulus that divides the divisor evenly. This process is described by (38) and in (40). Note that for each division of the DENOM value by one or more powers of a base modulus, the NUMER value is also divided by the same amount as implied in (46). Not shown by the detail of the state diagram is the need to subtract NUMER by a minimum value to make it evenly divisible by one or more powers of the base modulus dividing DENOM. This is like mixed-radix conversion. More about the logic required to perform this operation is described in the hardware section of this paper and may also be analyzed in the code.

While repeatedly dividing the DENOM and NUMER values, there may become a state where the DENOM is no longer evenly divisible by a base modulus, yet the modulus m=2 still has a valid digit. This is shown as the transition case: "**no zeros & pwr_valid_2 > 0**". The reason is that the DENOM value is incremented and will become evenly divisible by at least the base modulus m=2. However, if DENOM is not evenly divisible by a base modulus, and furthermore the base modulus m=2 is invalid, the case "**no zeros & pwr_valid_2 = 0**" is true and a base extension of the DENOM value is required. Additionally, the NUMER value will also be base extended in order that all moduli are returned to their *normalized* state. This is a priority transition in the state diagram. After base extension of both the DENOM and NUMER values, the FSM will transition back to the DIV_DENOM state if DENOM is evenly divisible by a base modulus or will transition back to the **INC_DENOM** state to increment the DENOM value and then transition back to the **DIV_DENOM** state.

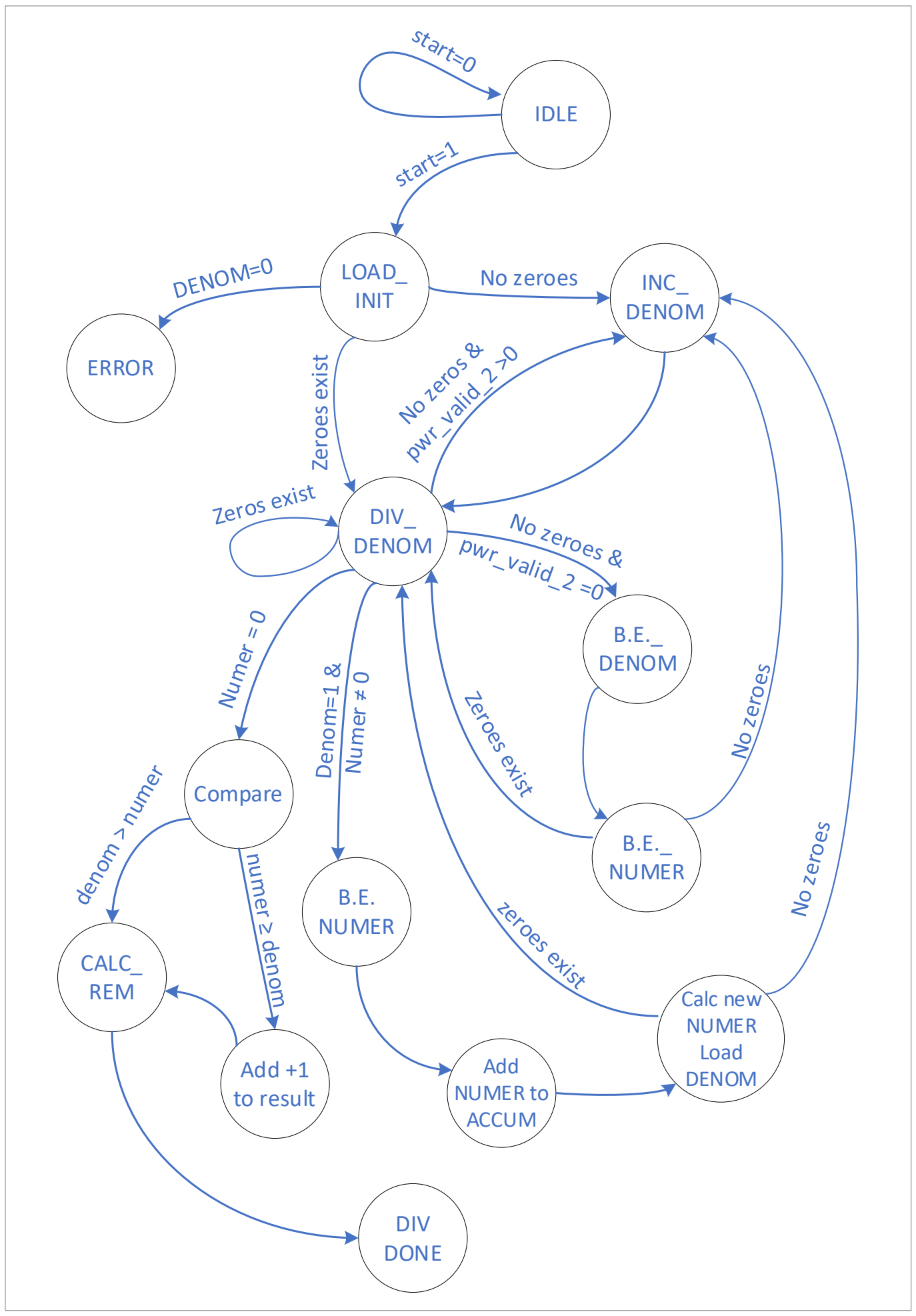

*Figure 5 - Simplified state transition for division algorithm*

After repeated divisions by various powers of base moduli, and if the NUMER value does not go to zero, the DENOM value is reduced to a value of one. In this case, the NUMER value is either the full quotient, or a portion of the quotient. This state transition is shown as the "**Denom=1 & Numer ≠ 0**" case which transitions the FSM to the **B.E. NUMER** state. In the **B.E. NUMER** state, the NUMER value is base extended since all digits need to be defined for the quotient result, or a portion of the quotient result. The NUMER value is then added to an accumulator which accumulates the quotient result. The summation of portions of the quotient is shown in (47).

The FSM machine then transitions to the state **Calc new NUMER Load DENOM** process to calculate a new NUMER value and reload a fresh copy of the divisor into DENOM. The new NUMER value follows the formula provided in (45), where the multiplication and subtraction are each PAC operations using modular arithmetic. Note that an alternate calculation is used in the implementation of the code. In the code, the original dividend value (X) is subtracted by the product of the current accumulated quotient value times the original divisor value (Y). This calculation retrieves the remaining portion of the numerator which has yet to be divided by the divisor. This shortcut reduces the number of RNS values that need to be stored and returns the same value as in (45). Once a new NUMER is ready, and the original divisor is loaded into DENOM, the FSM transitions to the **DIV_DENOM** state if DENOM is evenly divisible by a base modulus, or transitions to the **INC_DENOM** state to increment the denominator.

If during the reduction of DENOM and NUMER values in the **DIV_DENOM** state the NUMER value goes to zero, the FSM transitions to the state **Compare**. The **Compare** state compares the original NUMER value before reduction in the **DIV_DENOM** state to the full value of the divisor value. If the NUMER value is greater than the original divisor (DENOM before entering **DIV_DENOM**), then the quotient accumulator is incremented by one according to the case statement of (48). At this point, a complete quotient is contained in ACCUM according to (47) and (48). If the division algorithm shall also return a remainder, as does the code, the process of calculating the remainder is performed using PAC operations in the state **CALC_REM**. The full division is complete at this point.

## Divide Algorithm Example

To further illustrate the RNS integer division algorithm presented in this paper, a numerical example of an RNS integer division is tabulated in Table 3, Table 4, and Table 5. In these tables, the column labeled **State** corresponds to the state diagram of Figure 5, the column labeled **Register** denotes the RNS value stored in the hardware registers as discussed in the section entitled General RNS Processing Architecture. The digit labels $d_1$ *through* $d_8$ tabulate each digit value of the corresponding modulus. Note only power-based moduli are tabulated for convenience; note that when supported, non-power-based moduli affect the calculation during base extension, comparison, and undergo all multiplication by multiplicative inverses. However, non-power-based moduli are not dynamic and do not change. Moreover, in this example non-power-based moduli are not required, since the range of the RNS number system of only power-based moduli is more than enough to support the values in the example.

The **Notes** column presents the decimal equivalents of the RNS numbers, enhancing readability and making the example easier to follow. However, determining the value of an RNS number by inspection alone is not straightforward, which adds complexity to the development of RNS algorithms. To assist with this, the code materials include a link to the **RNS APAL code library**, a software tool designed to facilitate the study and analysis of RNS operations. This library offers convenient conversion between RNS and decimal representations and allows users to step through the division algorithm presented in this paper.

In Table 3, Step 0, the RNS values for dividend and divisor are loaded into the hardware registers NUMER and DENOM respectively. The ACCUM register is initialized to zero. The DENOM register employs logic to detect if the value is evenly divisible by one or more powers of a base modulus. In this case, there is no divisibility, so the divisor is incremented by one in Step 1; this makes the DENOM value evenly divisible by 256. In Step 2 the first part of the algorithm state INC_DENOM is executed, that is, the NUMER value is subtracted by a value that makes it also evenly divisible by 256. In Step 3, both NUMER and DENOM values are divided (scaled) by the full value of the modulus $M_5$ by multiplication of the multiplicative inverse of 256 with respect to each digit modulus. After the scaling, the modulus $M_5$ is invalidated as indicated by the asterisk as justified by (28). In Step 4, since the base modulus of two is now invalid, a base extension is performed to recover the full power and digit value for $M_5$ of both NUMER and DENOM registers. This is a critical step since without a valid base two modulus, there is no possibility to guarantee the DENOM value is divisible by at least the two modulus after being incremented.

After base extension in Step 4, the state machine transitions to INC_DENOM since after base extension the DENOM register is not divisible by any base modulus; therefore, DENOM is incremented in Step 5 where it now becomes divisible by the value 25, i.e., two powers of base modulus five. The algorithm state transition to DIV_DENOM in Step 6 where the NUMER register is subtracted by the value 24 to make it evenly divisible by 25. In Step 7, the NUMER and DENOM registers are divided (scaled) by two powers of the base modulus $m_2 = 5$. The effect of this scaling operation is to modify the digit value $d_2$ by direct division by 25 and to scale the modulus value to $M_2 = 5$ which is highlighted in blue. Note this reduction in the value of the digit and modulus is justified by (33). All other digits are scaled by (modular) multiplication with the inverse of 25 with respect to each digit's modulus. The multiplicative inverse of the value 25 with respect to each digit's modulus is highlighted and may also be ascertained using Table 6.

The process of repeated scaling is tabulated in Step 8 through Step 17. In these steps, the DENOM value is detected to be divisible by some power of a base modulus. In each scaling operation, the NUMER value is adjusted to be divisible by the same power of that base modulus, and the scaling operation is performed on both the NUMER and DENOM registers. Each time, the modulus being scaled is reduced, and the value of its digit is directly divided. As with the division by the value 256, it is possible that a digit modulus is completely invalidated. The remaining digits are scaled using modular multiplication by a multiplicative inverse.

In Step 17, the DENOM value is detected to equal one. This terminates the divisor decomposition sequence of (38). In the state transition diagram of Figure 5, the path denoted by the case DENOM=1 & NUMER≠0 is followed. This state transition indicates the NUMER value has not been divided out entirely. However, a close approximation to the quotient exists in the NUMER register. In Step 18, the NUMER register is base extended so that the value represented by the NUMER register is fully normalized. In other words, the original and full modulus for each digit is restored as highlighted in blue. This is important since the quotient result will be derived from a summation of the ACCUM register as described by (47). When summing portions of the quotient for a resulting quotient, the RNS value must be in a standard RNS representation defined by the machine word. In Step 19, the value of the NUMER register is added to the ACCUM register which was initialized to zero at Step 1 (not shown).

In Table 5, Step 20, a new NUMER value is calculated according to (45) so that a new partial quotient of (46) can be computed using repeated scaling. (Alternatively, the new NUMER value can be calculated using the original value of the dividend subtracted by the updated value of the ACCUM register times the original divisor.) Also, the value of the DENOM register is updated with a fresh copy of the divisor. In Table 5, Step 21 through 28, a repeated scaling of the NUMER value is illustrated. The DENOM value is not shown for brevity, however, the decomposition sequence for the DENOM value is the same shown in Step 1 through Step 17. In other words, the effective value for $\hat{Y}$ in (46) remains the same. The reader will note that once the decomposition of the DENOM is known, the division is focused on repeated scaling of the NUMER register.

| Step | State | Register | Digit index ($i$) | $d_1$ | $d_2$ | $d_3$ | $d_4$ | $d_5$ | $d_6$ | $d_7$ | $d_8$ | Notes |
|---|---|---|---|---|---|---|---|---|---|---|---|---|
| | | | Digit modulus | $M_1$ = 121 ($11^2$) | $M_2$ = 125 ($5^3$) | $M_3$ = 169 ($13^2$) | $M_4$ = 243 ($3^5$) | $M_5$ = 256 ($2^8$) | $M_6$ = 289 ($17^2$) | $M_7$ = 343 ($7^3$) | $M_8$ = 361 ($19^2$) | |
| 0 | LOAD_INIT | NUMER | Starting value | 49 | 71 | 69 | 18 | 177 | 0 | 227 | 197 | $x=987654321_{10}$ |
| | | DENOM | Starting value | 67 | 68 | 138 | 103 | 255 | 92 | 40 | 274 | Y=11634943 |
| 1 | INC_DENOM | DENOM | Increment DENOM | 68 | 69 | 139 | 104 | 0 | 93 | 41 | 275 | DENOM is now divisible by 256 |
| 2 | DIV_DENOM (part 1) | NUMER | Subtract 177 from NUMER | 114 | 19 | 61 | 84 | 0 | 112 | 50 | 20 | NUMER now divisible by 256 |
| 3 | DIV_DENOM (part 2) | | Value of $\left\|\frac{1}{256}\right\|_{M_i}$ | 26 | 21 | 68 | 187 | * | 35 | 205 | 55 | Multiplicative inverses |
| | | NUMER | Multiply by $\left\|\frac{1}{256}\right\|_{M_i}$ | 60 | 24 | 92 | 156 | * | 163 | 303 | 17 | 987654144/256= 3858024 |
| | | DENOM | Multiply by $\left\|\frac{1}{256}\right\|_{M_i}$ | 74 | 74 | 157 | 8 | * | 76 | 173 | 324 | 11634944/256 = 45449 |
| 4 | B.E. NUMER | NUMER | Base extend | 60 | 24 | 92 | 156 | 104 | 163 | 303 | 17 | Base extension to recover the modulus 256 |
| | B.E. DENOM | DENOM | Base extend | 74 | 74 | 157 | 8 | 137 | 76 | 173 | 324 | |
| 5 | INC_DENOM | DENOM | Increment DENOM | 75 | 75 | 158 | 9 | 138 | 77 | 174 | 325 | DENOM is now divisible by 25 |
| 6 | DIV_DENOM (part 1) | NUMER | Subtract 24 from NUMER | 36 | 0 | 68 | 132 | 80 | 139 | 279 | 354 | NUMER now divisible by 25 |
| 7 | DIV_DENOM (part 2) | | Value of $\left\|\frac{1}{25}\right\|_{M_i}$ | 92 | * | 142 | 175 | 41 | 185 | 247 | 130 | Multiplicative inverses |
| | | NUMER | Multiply by $\left\|\frac{1}{25}\right\|_{M_i}$ | 45 | 0 | 23 | 15 | 208 | 283 | 313 | 173 | 3858000/25= 154320 |
| | | DENOM | Multiply by $\left\|\frac{1}{25}\right\|_{M_i}$ | 3 | 3 | 128 | 117 | 26 | 84 | 103 | 13 | 45500/25= 1818 |
| | | | Modulus reduced | | $M_2$=5 | | | | | | | New modulus for $M_2$=5 |
| 8 | DIV_DENOM (part 1) | NUMER | Subtract 6 from NUMER | 39 | 4 | 17 | 9 | 202 | 277 | 307 | 167 | NUMER now divisible by 9 |
| 9 | DIV_DENOM (part 2) | | Value of $\left\|\frac{1}{9}\right\|_{M_i}$ | 27 | 4 | 94 | * | 57 | 257 | 305 | 321 | Multiplicative inverses |
| | | NUMER | Multiply by $\left\|\frac{1}{9}\right\|_{M_i}$ | 85 | 1 | 77 | 1 | 250 | 95 | 339 | 179 | 154314/9= 17146 |
| | | DENOM | Multiply by $\left\|\frac{1}{9}\right\|_{M_i}$ | 81 | 2 | 33 | 13 | 202 | 202 | 202 | 202 | 1818/9= 202 |
| | | | Modulus reduced | | | | $M_4$=27 | | | | | New modulus for $M_4$=27 |
| 10 | DIV_DENOM (part 1) | NUMER | Subtract 0 from NUMER | 85 | 1 | 77 | 1 | 250 | 95 | 339 | 179 | NUMER already divisble by 2 |
| 11 | DIV_DENOM (part 2) | | Value of $\left\|\frac{1}{2}\right\|_{M_i}$ | 61 | 3 | 85 | 14 | * | 145 | 172 | 181 | Multiplicative inverses |
| | | NUMER | Multiply by $\left\|\frac{1}{2}\right\|_{M_i}$ | 103 | 3 | 123 | 14 | 125 | 192 | 341 | 270 | 17146/2= 8573 |
| | | DENOM | Multiply by $\left\|\frac{1}{2}\right\|_{M_i}$ | 101 | 1 | 101 | 20 | 101 | 101 | 101 | 101 | 202/2= 101 |
| | | | Modulus reduced | | | | | $M_5$=128 | | | | New modulus for $M_5$=128 |

*Table 3 - Example RNS integer division step-by-step (part 1)*

| | | | Digit modulus | $M_1$ = 121 ($11^2$) | $M_2$ = 5 ($5^1$) | $M_3$ = 169 ($13^2$) | $M_4$ = 27 ($3^3$) | $M_5$ = 128 ($2^7$) | $M_6$ = 289 ($17^2$) | $M_7$ = 343 ($7^3$) | $M_8$ = 361 ($19^2$) | |
|---|---|---|---|---|---|---|---|---|---|---|---|---|
| Step | State | Register | Digit index (*i*) | $d_1$ | $d_2$ | $d_3$ | $d_4$ | $d_5$ | $d_6$ | $d_7$ | $d_8$ | Notes |
| 12 | DIV_DENOM (part 1) | NUMER | Subtract 2 from NUMER | 101 | 1 | 121 | 12 | 123 | 190 | 339 | 268 | NUMER now divisble by 3 |
| 13 | DIV_DENOM (part 2) | | Value of $\left\|\frac{1}{3}\right\|_{M_i}$ | 81 | 2 | 113 | * | 43 | 193 | 229 | 241 | Multiplicative inverses |
| | | NUMER | Multiply by $\left\|\frac{1}{3}\right\|_{M_i}$ | 74 | 2 | 153 | 4 | 41 | 256 | 113 | 330 | 8571/3 = 2857 |
| | | DENOM | Multiply by $\left\|\frac{1}{3}\right\|_{M_i}$ | 34 | 4 | 34 | 7 | 34 | 34 | 34 | 34 | 101/3= 34 |
| | | | Modulus reduced | | | | $M_4$=9 | | | | | New modulus for $M_4$=9 |
| 14 | DIV_DENOM (part 1) | NUMER | Subtract 1 from NUMER | 73 | 1 | 152 | 3 | 40 | 255 | 112 | 329 | NUMER now divisble by 2 |
| 15 | DIV_DENOM (part 2) | | Value of $\left\|\frac{1}{2}\right\|_{M_i}$ | 61 | 3 | 85 | 5 | * | 145 | 172 | 181 | Multiplicative inverses |
| | | NUMER | Multiply by $\left\|\frac{1}{2}\right\|_{M_i}$ | 97 | 3 | 76 | 6 | 20 | 272 | 56 | 345 | 2856/2= 1428 |
| | | DENOM | Multiply by $\left\|\frac{1}{2}\right\|_{M_i}$ | 17 | 2 | 17 | 17 | 17 | 17 | 17 | 17 | 34/2=17 |
| | | | Modulus reduced | | | | | $M_5$=64 | | | | New modulus for $M_5$=64 |
| 16 | DIV_DENOM (part 1) | NUMER | Subtract 0 from NUMER | 97 | 3 | 76 | 6 | 20 | 272 | 56 | 345 | NUMER is divisble by 17 |
| 17 | DIV_DENOM (part 2) | | Value of $\left\|\frac{1}{17}\right\|_{M_i}$ | 57 | 3 | 10 | 8 | 49 | * | 222 | 85 | Multiplicative inverses |
| | | NUMER | Multiply by $\left\|\frac{1}{17}\right\|_{M_i}$ | 84 | 4 | 84 | 3 | 20 | 16 | 84 | 84 | 1428/17= 84 |
| | | DENOM | Multiply by $\left\|\frac{1}{17}\right\|_{M_i}$ | 1 | 1 | 1 | 1 | 1 | 1 | 1 | 1 | 17/17 = 1 |
| | | | Modulus reduced | | | | | | $M_6$=17 | | | New modulus for $M_6$=17 |
| 18 | B.E. NUMER | NUMER | Base Extended | 84 | 84 | 84 | 84 | 84 | 84 | 84 | 84 | NUMER sum to accum |
| | | | Digit modulus | $M_1$ = ($11^2$) | $M_2$ = ($5^3$) | $M_3$ = ($13^2$) | $M_4$ = ($3^5$) | $M_5$ = ($2^8$) | $M_6$ = ($17^2$) | $M_7$ = ($7^3$) | $M_8$ = ($19^2$) | All NUMER Moduli restored |
| 19 | UPDATE_ ACCUM | ACCUM | NUMER added | 84 | 84 | 84 | 84 | 84 | 84 | 84 | 84 | NUMER sum to accum |

*Table 4 - Example integer RNS division step-by-step (part 2)*

| Step | State | Register | Digit index ($i$) | $d_1$ | $d_2$ | $d_3$ | $d_4$ | $d_5$ | $d_6$ | $d_7$ | $d_8$ | Notes |
|---|---|---|---|---|---|---|---|---|---|---|---|---|
| | | | Digit modulus | $M_1$ = 121 ($11^2$) | $M_2$ = 125 ($5^3$) | $M_3$ = 169 ($13^2$) | $M_4$ = 243 ($3^5$) | $M_5$ = 256 ($2^8$) | $M_6$ = 289 ($17^2$) | $M_7$ = 343 ($7^3$) | $M_8$ = 361 ($19^2$) | |
| 20 | CALC_NUMER | NUMER | 987654321-(84*11634943) | 108 | 109 | 138 | 114 | 5 | 75 | 297 | 285 | NUMER= 10319109 |
| 21 | DIV_DENOM | NUMER | Subtract 5 from NUMER | 103 | 104 | 133 | 109 | 0 | 70 | 292 | 280 | NUMER is divisble by 256 |
| | | NUMER | Multiply by $\left\|\frac{1}{256}\right\|_{M_i}$ | 16 | 59 | 87 | 214 | * | 138 | 178 | 238 | 10319109/256= 40309 |
| 22 | B.E. NUMER | NUMER | Base Extended | 16 | 59 | 87 | 214 | 117 | 138 | 178 | 238 | NUMER is normalized |
| 23 | DIV_DENOM | NUMER | Subtract 9 from NUMER | 7 | 50 | 78 | 205 | 108 | 129 | 169 | 229 | 40309-9= 40300 |
| | | NUMER | Multiply by $\left\|\frac{1}{25}\right\|_{M_i}$ | 39 | 2 | 91 | 154 | 76 | 167 | 240 | 168 | 40300/25= 1612 |
| | | | Modulus reduced | | $M_2$=5 | | | | | | | New modulus for $M_2$=5 |
| 24 | DIV_DENOM | NUMER | Subtract 1 from NUMER | 38 | 1 | 90 | 153 | 75 | 166 | 239 | 167 | 1612-1= 1611 |
| | | NUMER | Multiply by $\left\|\frac{1}{9}\right\|_{M_i}$ | 58 | 4 | 10 | 17 | 179 | 179 | 179 | 179 | 1611/9= 179 |
| | | | Modulus reduced | | | | $M_4$=27 | | | | | New modulus for $M_4$=27 |
| 25 | DIV_DENOM | NUMER | Subtract 1 from NUMER | 57 | 3 | 9 | 16 | 178 | 178 | 178 | 178 | 179-1= 178 |
| | | NUMER | Multiply by $\left\|\frac{1}{2}\right\|_{M_i}$ | 89 | 4 | 89 | 8 | 89 | 89 | 89 | 89 | 178/2= 89 |
| | | | Modulus reduced | | | | | $M_5$=128 | | | | New modulus for $M_5$=128 |
| 26 | DIV_DENOM | NUMER | Subtract 2 from NUMER | 87 | 2 | 87 | 6 | 87 | 87 | 87 | 87 | 89-2= 87 |
| | | NUMER | Multiply by $\left\|\frac{1}{3}\right\|_{M_i}$ | 29 | 4 | 29 | 2 | 29 | 29 | 29 | 29 | 87/3= 29 |
| | | | Modulus reduced | | | | $M_4$=9 | | | | | New modulus for $M_4$=9 |
| 27 | DIV_DENOM | NUMER | Subtract 1 from NUMER | 28 | 3 | 28 | 1 | 28 | 28 | 28 | 28 | 29-1= 28 |
| | | NUMER | Multiply by $\left\|\frac{1}{2}\right\|_{M_i}$ | 14 | 4 | 14 | 5 | 14 | 14 | 14 | 14 | 28/2= 14 |
| | | | Modulus reduced | | | | | $M_5$=64 | | | | New modulus for $M_5$=64 |
| 28 | DIV_DENOM | NUMER | Subtract 14 from NUMER | 0 | 0 | 0 | 0 | 0 | 0 | 0 | 0 | 14-14= 0 |
| | | NUMER | Zero detected | 0 | 0 | 0 | 0 | 0 | 0 | 0 | 0 | NUMER = 0 detected |
| 29 | COMPARE | OLD NUMER | Compare NUMER to DENOM | 108 | 109 | 138 | 114 | 5 | 75 | 297 | 285 | DENOM > NUMER (11634943 > 10319109) |
| | | DENOM | | 67 | 68 | 138 | 103 | 255 | 92 | 40 | 274 | |
| 30 | CALC_REM | REM | Y-A*X | 108 | 109 | 138 | 114 | 5 | 75 | 297 | 285 | REM=Y-(A*X) |
| 31 | DONE | QUOTIENT | | 84 | 84 | 84 | 84 | 84 | 84 | 84 | 84 | Move Accum To QUOTIENT |

*Table 5 - Example RNS integer division step-by-step (part 3)*

In Step 28, the NUMER register is equal to zero. Therefore, the path NUMER=0 in the state transition diagram is taken to the Compare state from the DIV_DENOM state. In this case, a comparison test according to (48) is performed. In the case of the example values presented, the last calculated NUMER value is less than DENOM (the full divisor). Therefore, there is no increment (add one) operation performed on the ACCUM register. Therefore, the value in the ACCUM register represents the complete quotient of the division, which is equal to 84. As a final operation, the remainder of the division is calculated using the value of the ACCUM register. This operation is purely a PAC operation and completes in only a few arithmetic cycles.

# General RNS Processing Architecture

The following sections describe the hardware organization used to realize the proposed division method and report initial implementation results. The architecture partitions RNS digit registers into distinct digit processing units (DPUs), following the general RNS hardware approach introduced in [9].

## Basic RNS Register

RNS numbers are composed of a series of RNS digits, as previously discussed. In hardware, these numbers are stored in registers, with each register containing multiple RNS digits. Each individual RNS digit is encoded in binary and stored in its own register as shown in Figure 6.

The largest digit modulus in the RNS word format determines the required bus width for interconnecting all digits to a shared data bus, referred to as the crossbar. To simplify hardware design, the digit bus width is typically standardized, with all digits encoded in fixed-width binary registers. For digits with moduli that require fewer bits than the fixed width, the unused most significant bits are set to zero. This approach enables the RNS value to be stored and processed in hardware as a binary-coded residue (BCR) number.

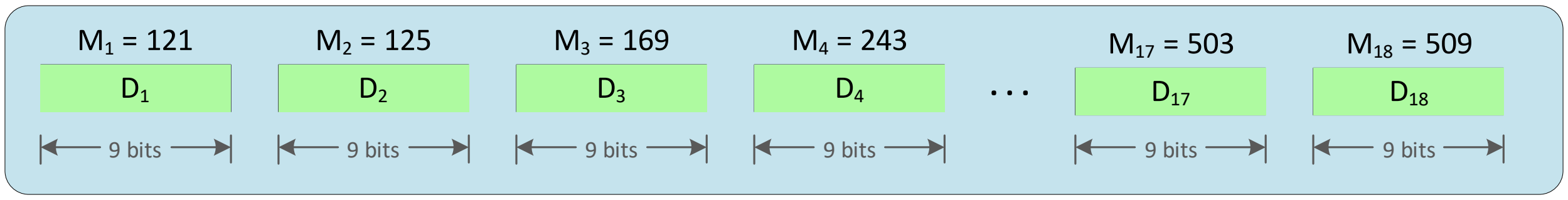

*Figure 6 - RNS word register comprises eighteen RNS digit registers.*

## Basic RNS Digit Control

During division, two RNS values are operated upon by a control circuit; an RNS dividend and an RNS divisor. For the multi-cycle division algorithm presented herein, both the dividend and divisor are processed in a digit-by-digit manner. Furthermore, each digit targeted for processing will result in an action that simultaneously affects all other digits. Figure 7 is provided to help the reader gain an understanding of the basic processor architecture for performing digit-by-digit operations in RNS.

In Figure 5, the shaded blue box represents an RNS processor that holds a complete RNS value. The processor consists of multiple DPUs, each containing an RNS digit register. These DPUs are connected to the divide register control unit, which can issue operations to all DPUs simultaneously via a bus labeled "op_code," shown in red. Also in red is the digit modulus select bus, labeled "mod_select," which determines the specific digit modulus or DPU being operated on. It is important to note that the selected DPU generates actions that influence all other DPUs. Throughout the discussion of DPU designs, we will often refer to the selected DPU as the selected modulus.

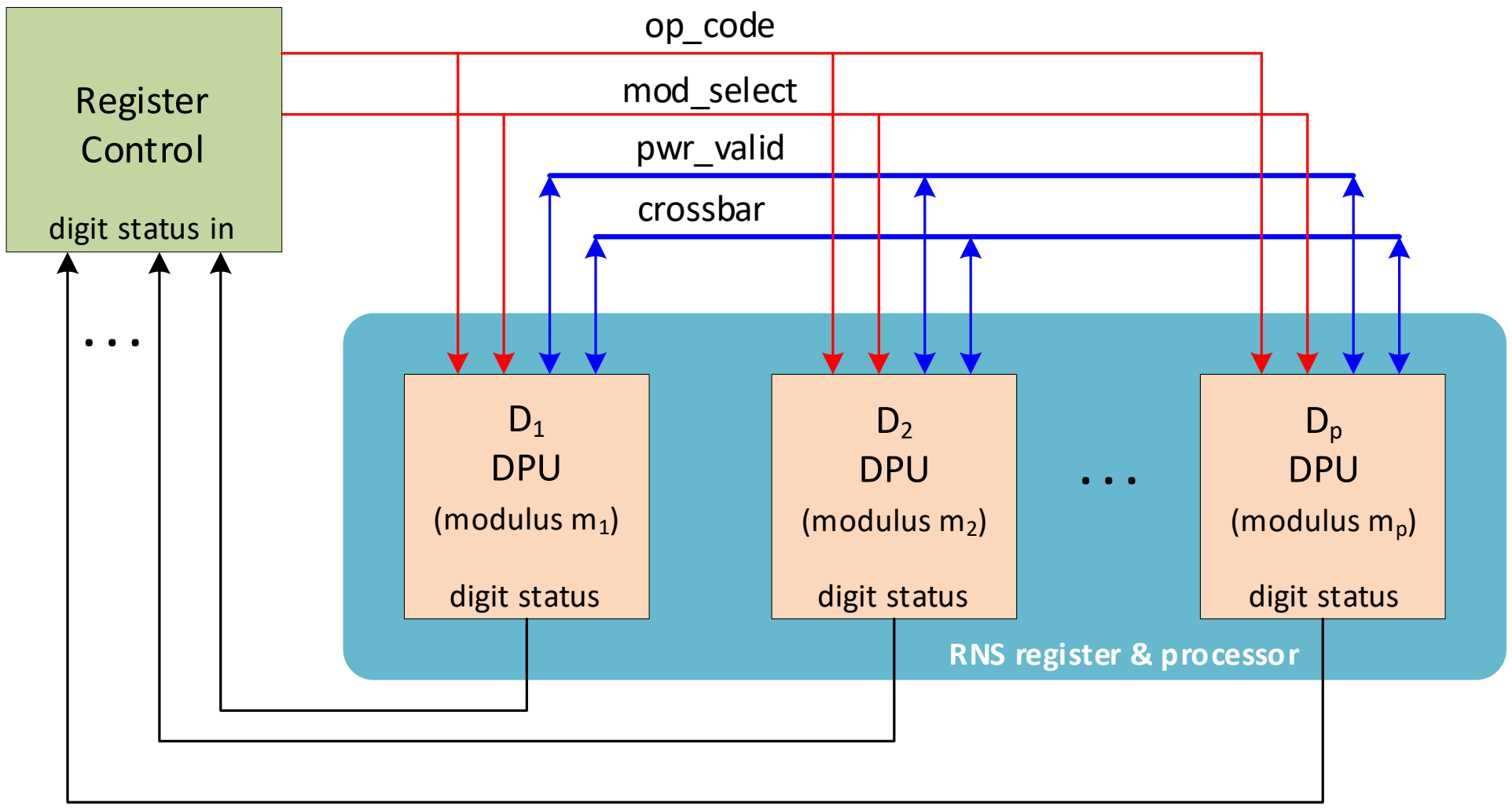

*Figure 7 – Basic RNS register processing unit and control.*

Two additional buses shown in blue are driven by tri-state drivers and will be referred to as tri-state buses. The tri-state bus allows data to flow from any selected DPU to all other DPUs. One tri-state bus, called the "*cross_digit*" bus, allows a selected DPU to share the value of its digit to all other DPUs. The second tri-state bus is called the "*pwr_valid*" bus, which allows the selected DPU to share the number of powers remaining in its modulus to all other DPUs.

Each digit processing unit of Figure 7 also outputs a set of signals that represent the state of the digit. One such signal is the *digit=0* signal, which informs the control unit that the digit value of a particular DPU is zero. Another control signal indicates if the digit is valid, herein called a *skip_digit* signal, which if set indicates the RNS digit is not valid. As mentioned earlier, digits may become invalid if their modulus value has been divided out from the register value. In typical operation, the control unit will sense the state of each DPU and execute a control sequence for division according to the evolving state of the RNS value.

## Digit Processing Unit for Mixed-Radix Conversion

The most basic digit-by-digit process is the mixed-radix conversion. The mixed-radix conversion is used for two purposes: 1) to perform base extension to recover the value of one or more invalid digit modulus or its powers, and 2) to perform comparison of two RNS values.

It should be noted that while the mixed-radix conversion produces a mixed-radix number, the mixed-radix number itself is not used. Instead, as each mixed-radix digit is generated it is processed then discarded. In this manner, the mixed-radix number is not stored; instead, as each mixed-radix digit is generated, it affects and advances the state of the divide; each mixed-radix digit is consumed before another mixed-radix digit is generated. Mixed-radix conversion is a mathematical computation allowing the division routine the ability to directly compare RNS values and recover invalid RNS digits.

### Basic Operation

A block diagram for a digit processing unit capable of mixed-radix conversion is shown in Figure 8. The digit processing block supports a modular subtractor shown in blue to provide the subtraction step of mixed-radix conversion. Note the subtrahend of the subtractor is connected to the crossbar bus; this provides a means to subtract a selected digit (from a different digit processing unit) from the digit value register shown in green. This subtraction will occur for all digits in the RNS register simultaneously. The digit processing unit of Figure 8 supports a register to store the subtraction result on a first clock cycle.

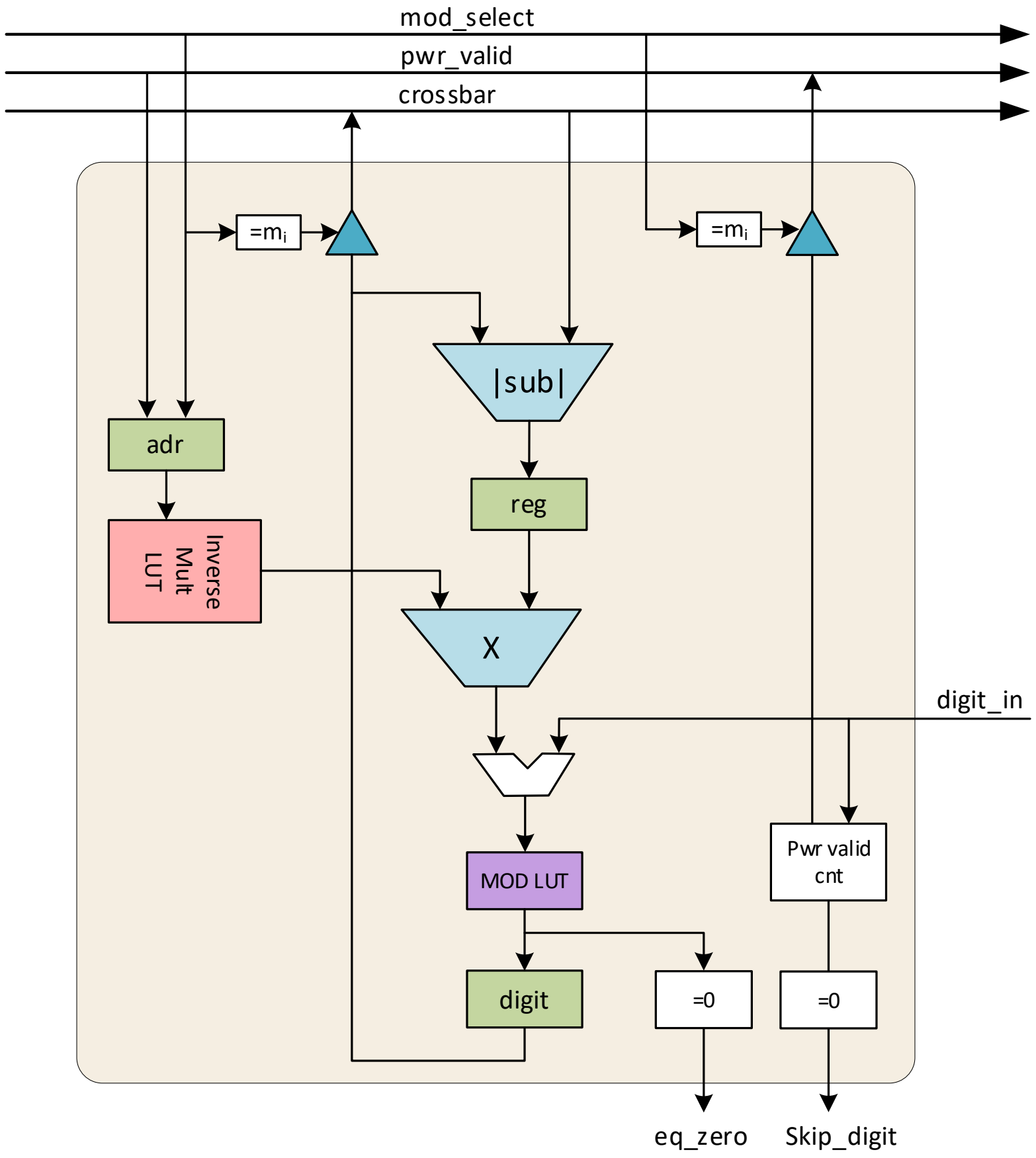

*Figure 8 - Basic digit processing unit for mixed-radix conversion.*

In Figure 8, a multiplier, also shown in blue, provides the multiply step of mixed-radix conversion, i.e., to multiply the result of the subtractor by a multiplicative inverse. This occurs on the second clock cycle; thus, this design requires two-clock cycles for every mixed-radix digit converted. The multiplicative inverses for each digit operation are stored using a look-up table (LUT) shown in red. Note the address of the inverse multiplier LUT, shown as a green address register, is provided by the concatenation of the *mod_select* and *pwr_valid* buses. This arrangement provides a means to index the correct inverse multiplier constant from the LUT, depending on what modulus is currently selected and the magnitude of its power.

The multiplier shown is a standard 9-bit binary multiplier supported by most FPGAs; however, for RNS arithmetic we must support a modular multiplier. Hence, we support a MOD function using a small set of LUTs as illustrated by the purple *MOD* block. The MOD function translates the output of the multiplier into a modular result, effectively creating a modular multiplier. Therefore, each digit processing unit will support a unique MOD LUT since the modulus, $m_i$, of each digit differs.

Tri-state buffers, shown as blue triangles, are supported to drive the tri-state buses; they are enabled if the *mod_select* bus value equals the digit processing unit's digit index *i*. If enabled, the value of the digit register is placed on the *crossbar* bus, and the value of the power valid count is placed on the *pwr_valid* bus. This happens when the digit modulus associated with the digit processing unit is selected for processing. Also shown is a *digit_in* input to load the RNS digit into the digit processing unit at initialization. As shown, a two-input mux is used to select the digit input during loading. Other features include a digit zero-check logic block and a power valid counter. The digit zero-check logic sends an indication to the controller when the digit is equal to zero. The power valid counter reflects the number of valid powers supported by the digit modulus. If the power valid count goes to zero, this is detected by the zero-check block which signals the controller the digit is invalid or *skipped*.

## Divisor Digit Processing Unit

The DPU for the RNS divisor (DENOM) will support three more operations in addition to mixed-radix conversion. These operations are: 1) the division decomposition sequence, and 2) arbitrary scaling by one or more powers of its base module; 3) support for a variable power modulus. A block design for a divisor digit processing unit is shown in Figure 9.

As seen in Figure 9, the structure of the divisor DPU continues to support a mixed-radix conversion capability as previously discussed. To support the division decomposition sequence and arbitrary scaling, two additional functions are required: 1) a function to increment the divisor, and 2) a function to divide the digit by one or more of its powers. To increment the divisor, an increment function shown in blue is provided to add one to the digit value and feed it back to the digit register. Note that after incrementing the digit, the resulting value is passed through the MOD function, since the addition must be modular. A 3-input mux is provided to steer the modular increment to the digit register.

To perform the divisor decomposition sequence and scaling operations, a *selected digit* must be divided by one or more of its powers. To perform this operation the **Div LUT** shown in blue is provided. Note that during the divide process, only one divisor digit is targeted for pure division; all other digits will undergo scaling using an inverse multiply. Note for the digit divide operation, the selected digit is known beforehand to be *evenly divisible* by one of more of its modulus, since the controller has selected the digit based on the availability of a "zero" using the *any_zero* status signal. Therefore, the result of the divide is a whole number digit value.

The DIV LUT divides the selected digit value by a constant; this constant value is a power of the base modulus of the selected digit. The number of constants supported is a design decision, which dictates how many modulus powers can be divided out in each scaling cycle. For example, it is possible to design a power-based DPU that allows only a single power to be divided per scaling cycle. The design shown herein allows any number of powers of a base modulus to be divided out of the selected digit modulus.

The digit processing unit for power-based digits must support a *variable modulus*. For example, for the modulus=125, there are three powers of the base modulus=5. Therefore, the DPU must support three different moduli, which are 125, 25, and 5. To support a variable digit modulus, a **MOD(pwr)** function is provided, shown as a blue block of Figure 9. The **MOD(pwr)** function changes the DPU's modulus based on the value of the power valid counter, shown as the block labeled ***pwr_valid_cntr***.

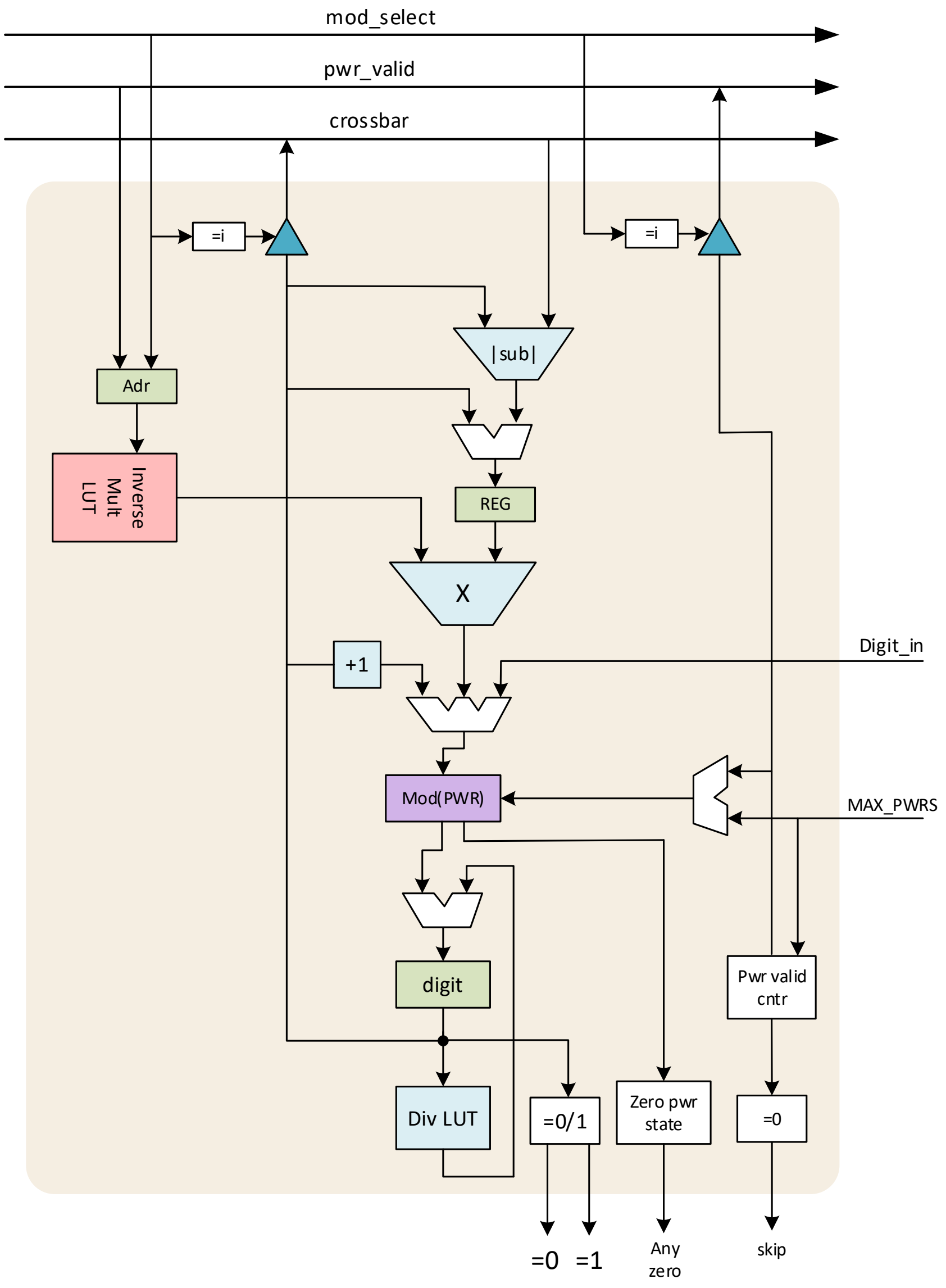

*Figure 9 - Digit processing unit for power-based digits of the divisor*

To implement the **MOD(pwr)** function, several sets of LUTs are supported, one set for each power supported. During operation, one of the LUT sets is selected by the value of the power valid counter. There are other solutions for the variable MOD function not discussed here that are candidates for future investigation.

Another important feature for the divisor DPU is the so-called "*any zero*" detection. This signal is true if the digit value is evenly divisible by one or more powers of its base modulus. The naming convention is derived from the observation that if a power-based modulus is represented using a fixed-radix number, where the radix is the value of the base modulus, a value that is evenly divisible by one or more of its powers will have zeros as the last digits. For example, given the binary modulus 256 and a digit value of $192_{10}$=$1100_0000_2$, note the binary representation has zeroes for the last 6 digits of its binary (fixed radix) representation. Therefore, we know the value 192 is evenly divisible by 6 powers of the base modulus=2, which is the value $2^6$=64.

The **MOD(pwr)** function can detect if any power of the base modulus is "zero", and if so, the "*any_zero*" signal is latched by the ***zero_pwr state*** block as shown in Figure 9. The divide controller monitors the *any_zero* signal from all power-based digits to determine which digit to select for scaling the denominator. Since the digit modulus of power-based digits can vary, it is important to notify all other DPUs of the current state of the selected digit's modulus. This is important because the multiplicative inverse for different powers of a modulus is generally different. When a digit is selected using the *mod_select* control, a tri-state buffer transmits the value of its power valid counter via the *pwr_valid* bus to all other DPUs.

Lastly, in addition to the detection of zero, the controller for the divisor DPU's will require that an RNS value =1 is detected (all valid digits equal one). The reason is that repeated division during divisor decomposition results in a terminating value of one. Note also that during a digit load, the ***pwr_valid_cntr*** must be loaded with the maximum powers of the power-based modulus.

## Dividend Digit Processing Unit

The *dividend* DPU for power-based modulus must support a similar set of functions as a power-based *divisor* DPU but is organized to achieve the required operations on the dividend RNS value. A block design for a dividend DPU is shown in Figure 10.

The DPU for the dividend value does not support an increment function. This is only needed for the divisor. However, the dividend does support scaling by an arbitrary power modulus, even if the digit of the selected modulus is not evenly divisible by the power modulus. To do this, the dividend supports a similar process used in mixed-radix conversion; it subtracts the digit by a value that makes it evenly divisible by the power modulus. All other dividend digits are also subtracted by that value. In contrast, during mixed-radix conversion, the entire digit value is subtracted from the selected modulus, and hence every other digit modulus via the crossbar, to make the RNS value evenly divisible by the selected modulus.

For the dividend DPU, subtracting the digit of the selected modulus to make it evenly divisible by some power of the modulus is more complex than for mixed-radix conversion. To do this, the dividend DPU will keep an offset value for each supported power of the divisor DPU of the same modulus. For each offset value, its assigned value is the **MOD(pwr)** function. For example, if all three powers of the base modulus 5 are supported for division, the dividend DPU will latch and maintain three offset values, one for each of the powers 125, 25 and 5. If the divisor DPU divides by the power 25, the dividend will subtract the current digit $d_i$ by the value $|d_i|_{25}$. To further the example, if $d_i$ = 93, the value $|93|_{25}$=18; then subtracting: 93-18=50 which is evenly divisible by the power 25.

The dividend DPU also supports a full digit register to make the subtraction circuitry simpler. In other words, the full RNS digit is a value based on the current value of the **pwr_valid_cntr**, since this counter defines the current modulus power, which is the largest power supported by the digit modulus in the current state. The full digit undergoes subtraction by one of the digit offsets based on the value of the *pwr_valid* bus which is driven by the divisor digit during scaling. The *pwr_valid* bus will contain the power of the modulus that divides the divisor evenly. The **pwr_valid_cntr** is decreased by the number of powers indicated by the *pwr_valid* bus since the digit modulus changes because of the digit operation.

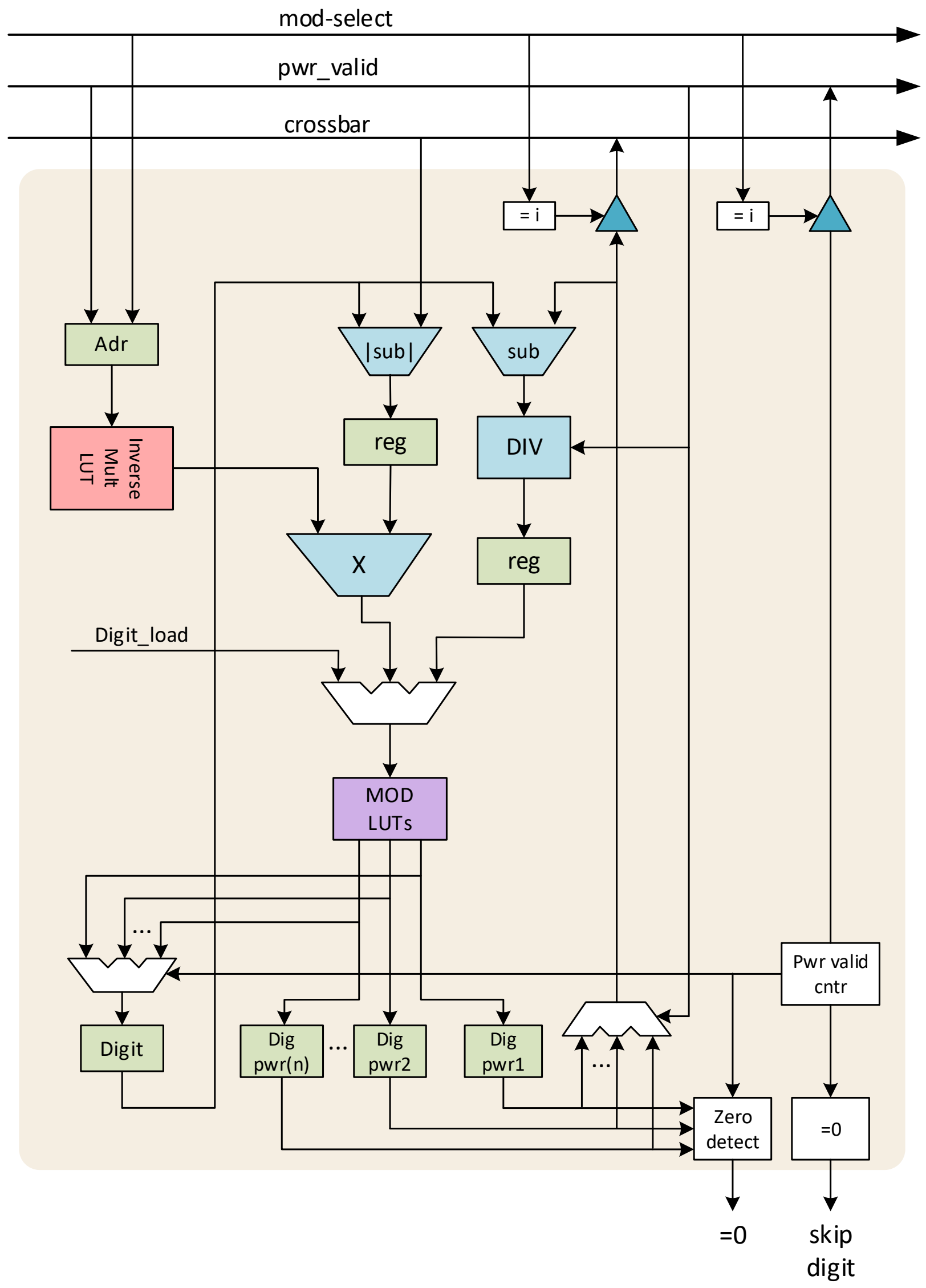

*Figure 10 - Dividend DPU for a power-based modulus.*

For the selected dividend digit, once it is made divisible by some power of the selected modulus by subtraction, it undergoes direct division as shown by the blue DIV block. The DIV block is identical to the DIV block of the divisor. It is comprised of LUTs to divide by one of the supported powers; the design of the DIV block can take advantage of the fact that the digit is known to be evenly divisible by a constant to reduce resources. All other DPUs will also subtract the dividend offset value since this value is gated to the crossbar by the selected DPU digit. However, for un-selected DPUs, they will undergo *scaling* by using inverse multiplication, which is valid since the modulus are coprime.

Note that the modulus of the divisor and the modulus of the dividend track each other on a digit-by-digit basis exactly. This is always the case. Therefore, the **pwr_valid_cntr** block can be shared between the DPU of the divisor and dividend for each digit modulus. The diagram of Figure 10 shows a duplicated **pwr_valid_cntr** for sake of clarity; alternatively, the counter can be repeated in hardware as a design decision to reduce routing delay.  Also note two status signals are available to inform the controller that 1) the digit is zero and 2) the digit is invalid or skipped because the **pwr_valid_cntr** has reached zero.

## Non-Power-Based DPU

A non-power-based version of the divisor DPU and dividend DPU is required if not all digits are power-based since these DPUs do not need to support a variable modulus.  It is possible to scale by these moduli, but typically a controller will not do so by design choice since the probability the digit is zero is small.  For example, for the digit modulus=509, when equal to zero will allow the RNS value to be divided by 509.  However, the chance of the digit being equal to zero is 1/509 on average, and therefore a designer might not deem it important to support action on this condition.  This design choice follows the idea that it is the small base moduli of the power-based digits that are frequently equal to zero, and these values form the basis for the divide algorithm.

In general, a designer will strip off un-needed functions required by power-based DPUs to support the non-power-based DPUs.  This means the non-power-based DPUs are like the basic DPU for mixed-radix conversion.

## Recombination Digit Processing Unit

Perhaps the most difficult operation to support in RNS arithmetic is that of base extension, that is, recovering the value of an invalid (skipped) digit given the value of all other valid (non-skipped) digits; the value of non-skipped digits is the current value of the RNS number.  Base extension is critical, since the ability to continue to use a digit of a specific modulus depends on recovering the digit's value back to a valid state after it has been (completely) divided out in a previous scaling operation.

In many papers, it is proposed to use the Chinese remainder theorem (CRT) to recover invalid digits.  However, circuits based on the Chinese remainder theorem rely on the existence of a redundant RNS digit; unfortunately, the existence of a redundant digit is not always maintained or guaranteed during general-purpose processing in RNS.  Therefore, the design herein illustrates how mixed-radix conversion is used to recover an invalid digit.  Furthermore, mixed-radix conversion can base extend not only one digit, but multiple invalid digits in the same operation.  Interestingly, the longer base extension is delayed, the more digits are recovered in one operation, and the more efficient the base extend operation becomes.

In a previous section, this paper introduced the idea of mixed-radix conversion where the generated mixed-radix digits are used immediately and then discarded.  This technique is also used to recover RNS digits by recombining the mixed-radix digits directly back to their RNS representation.  To perform this operation, a new digit processing unit is introduced which is termed a recombination unit in this paper.  To simplify the explanation of its operation, we first present the recombination unit of Figure 11.

To understand the recombination unit of Figure 11, it is noted the mixed-radix number format is a weighted and positional number system.  Like a fixed-radix number, each mixed-radix digit represents a portion of the number's value, i.e., that portion equals a weight (digit value) multiplied by the powers of all previous least significant digits.  This can be seen in (35).

The order of mixed-radix processing may vary.  Moreover, whether a particular RNS digit modulus is valid also determines if that digit will be included in the mixed-radix conversion.  To make matters a bit more complex, the mixed-radix conversion must handle a digit modulus that will vary, since its power may have been reduced by a scaling operation.  Therefore, the modulus and digit order can change from one mixed-radix conversion to another.

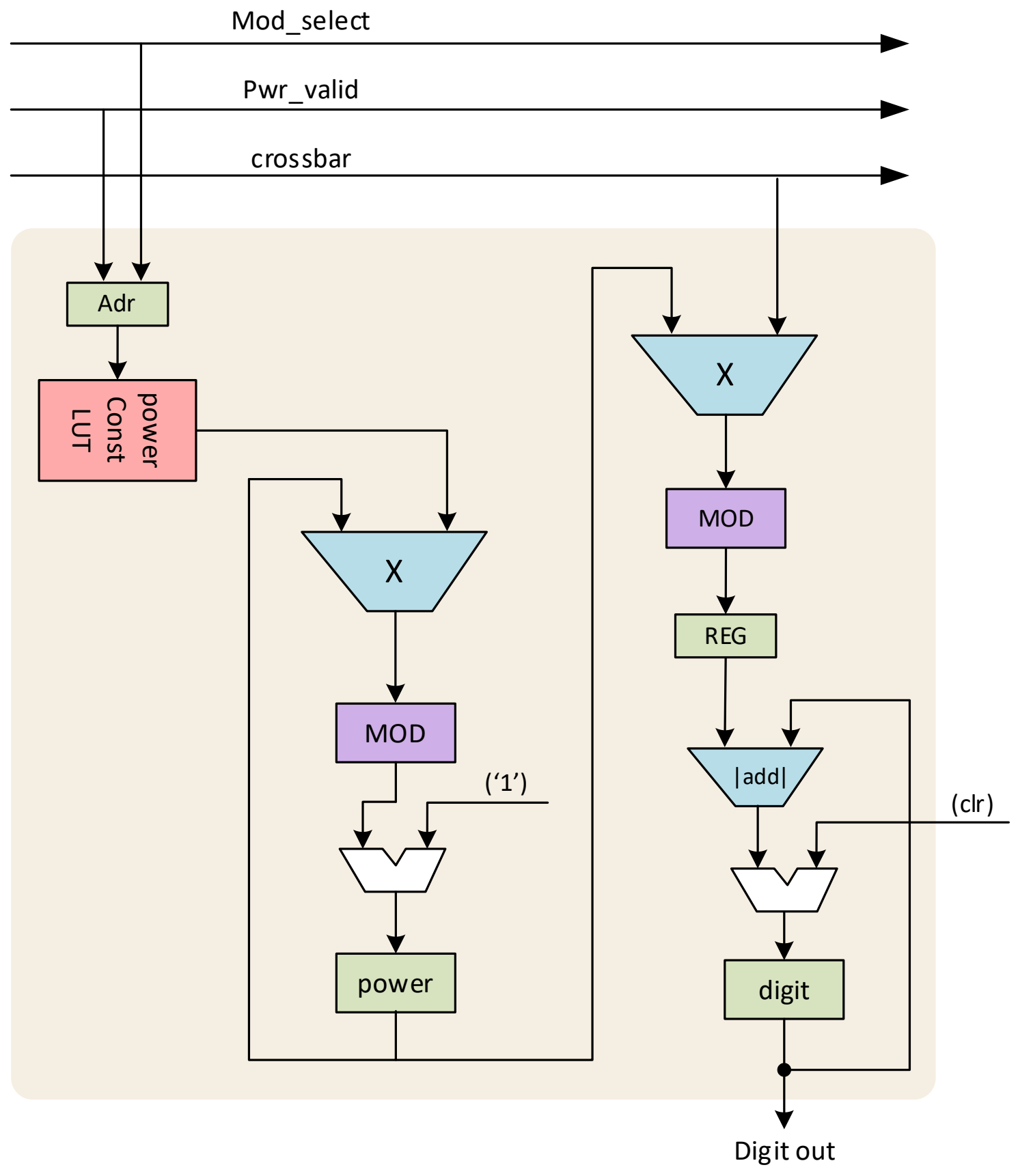

*Figure 11 - Simplified digit recombination unit.*

To address these complexities, the digit recombination of Figure 11 will support a digit accumulator register labeled **digit** and a digit power register labeled **power**.  Once again, the digit register will contain a modular value, the values stored in these registers will not exceed the fixed digit width of the RNS system.  In the first step of base extension, the recombination accumulator (digit register) is cleared while the power register is set to one.  As shown in Figure 11, the power register starts with number one, i.e., the power of the first mixed-radix digit.  On cycle 2 of each digit processing cycle, the power register is multiplied by the modulus of the processed digit using modular multiplication.  In this way, each power register of each recombination DPU maintains the power of all previously processed mixed-radix digits.

As mixed-radix digits are generated, they are transmitted on the crossbar of Figure 11.  On cycle 1 of each digit processing cycle, the mixed-radix digit is multiplied by the current power value and accumulated into the digit register of the recombination unit, again using modular multiplication.  In this way, the contribution of each weighted digit of the mixed-radix conversion is added to the RNS digit accumulator. This process is repeated in each digit recombination unit simultaneously.

The order of mixed-radix conversion is chosen by the controller of Figure 7. Typically, a left to right order of digits is chosen by the controller for mixed-radix conversion, but only valid digits will be processed. Any invalid RNS digit is marked as skipped and will be recovered in the digit register of the recombination unit; all non-skipped digits will be processed by the mixed-radix conversion performed in the divisor digit processing units and numerator digit processing units.

For simplicity of hardware organization, it is typical that even non-skipped power-based digits are recovered using a recombination digit unit, since there is a requirement that each power-based digit be fully restored to its maximum power modulus. Furthermore, non-skipped digits with their full power modulus, including non-power-based digits, can be base extended. This is an optional and redundant operation, since the fully recovered value of a valid digit is the same value. To save power, this redundant operation can be avoided by simply latching the value of the non-skipped, full power modulus. Non-power-based digits are always valid prior to mixed-radix conversion; their value may enter mixed-radix conversion, but likewise there is no need to support a digit recombination unit for these digits.

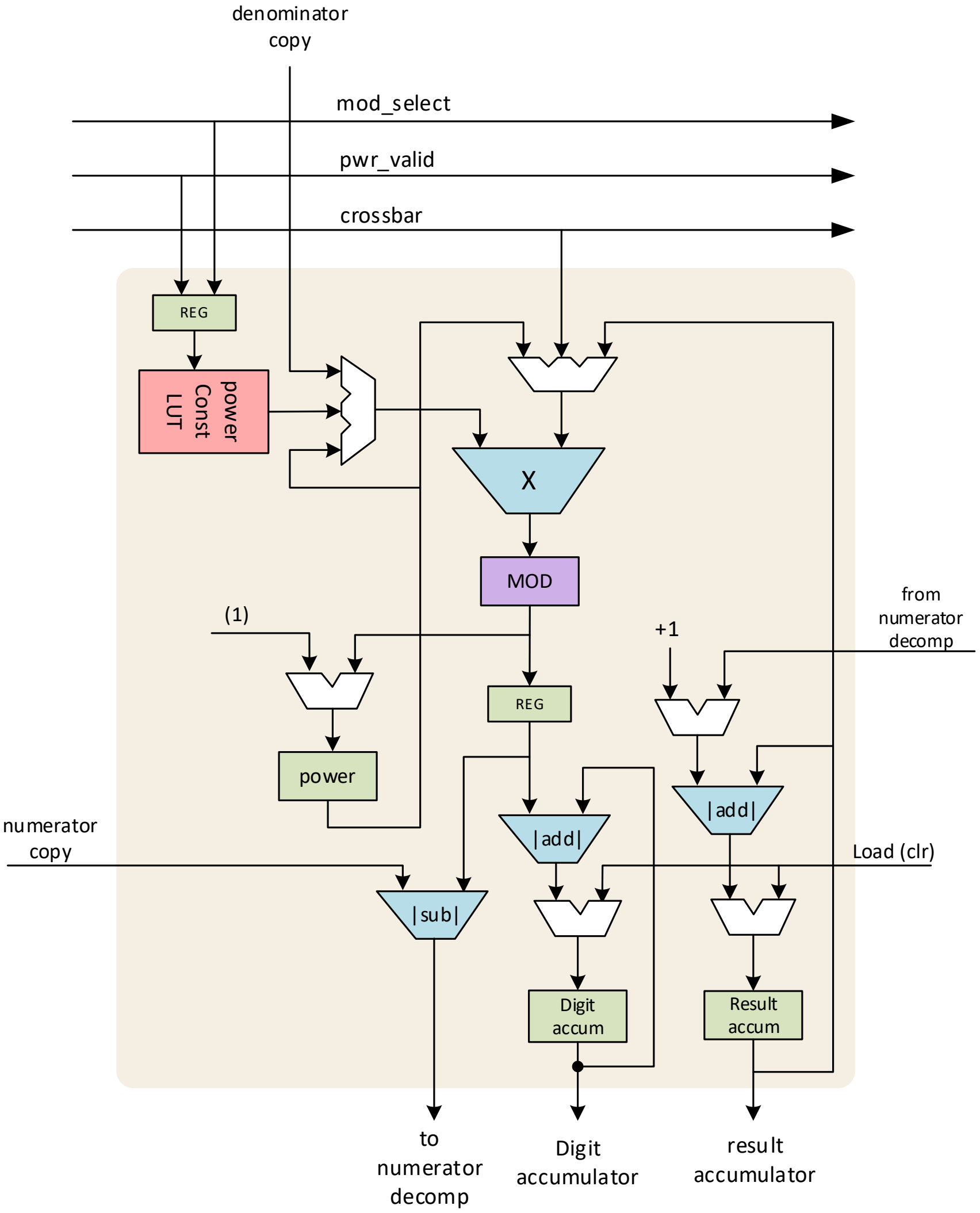

*Figure 12 - Recombination digit processing unit with shared multiplier, arithmetic unit and result accumulator.*

The divide algorithm also requires subtraction and multiplication of RNS numbers for each iteration as shown in equation (45). These RNS integer operations occur without digit to digit carry and are performed in a single clock cycle. In the design of this paper, these operations are integrated into the recombination unit as a convenient design choice. Also note that since the power multiply and the mixed-radix digit multiply occur on different cycles of each digit operation, it is possible to share a single modular multiplier. Thus, a final design for the digit recombination unit is shown in Figure 12.

## Summary of DPU Types

The number and type of DPU's for an integer division unit may vary. However, the general requirements are that power-based DPU's are employed for low value prime base moduli: 2, 3, 5, 7, etc. The theoretical minimum number of base moduli to guarantee convergence is an open problem of interest, however, the base modulus=2 is required for the algorithm as described. To extend the range of the RNS integer divide, any number of additional non-power-based DPU's are employed. Finally, a power-based recombination DPU with arithmetic capability is provided for each power-based modulus. An arithmetic DPU is provided for each non-power-based modulus. The overall organization comprises three RNS register processors, as illustrated in Figure 13.

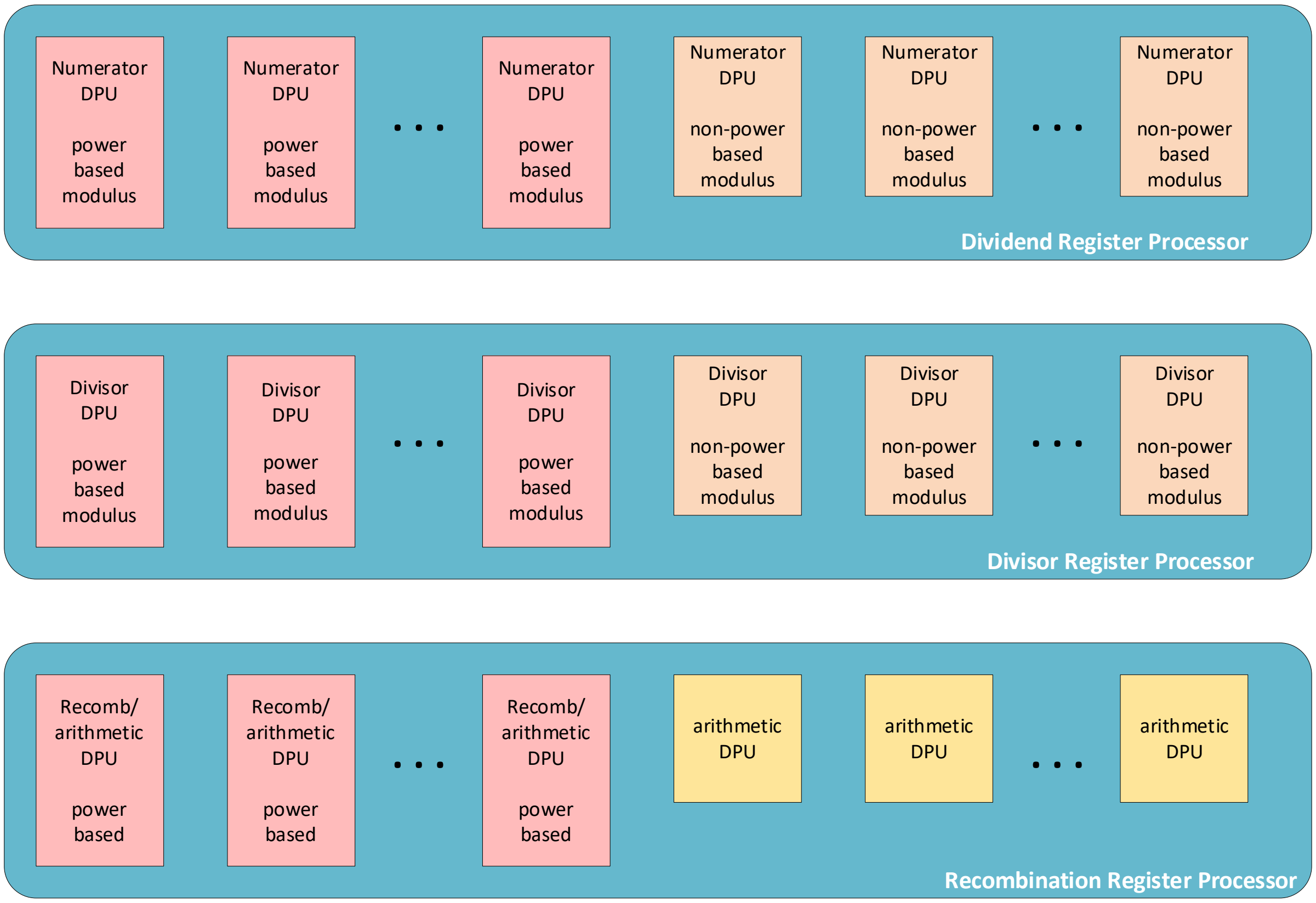

*Figure 13 - Organization of three RNS register processors and their DPUs.*

For naming conventions used in the preceding sections, the Dividend Register Processor of Figure 13 is referred to as the NUMER register, the Divisor Register Processor is referred to as DENOM register and the Recombination Register is referred to as RECOMP register. The ACCUM register and digit result registers are contained in the RECOMP register.

## Register Processor Module Connections

The three required RNS register processors are interconnected to a central control unit. The overall connections are illustrated in Figure 14.

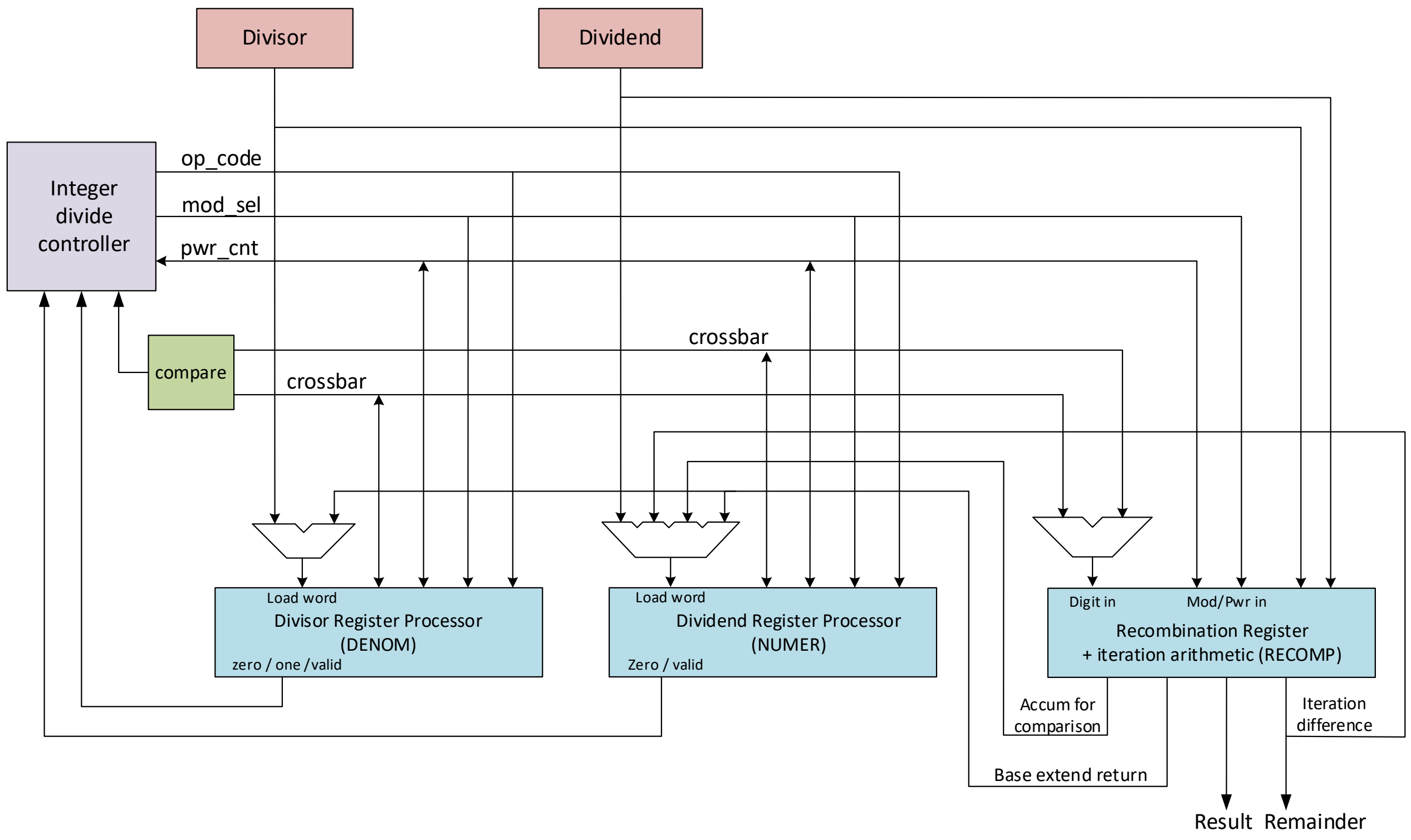

*Figure 14 - Inter-connections of RNS register processing units.*

In Figure 14, the initial RNS dividend and divisor, shown in red, are loaded into their respective blue register/processor modules. Status from these processors is routed to the integer divide controller, shown in purple, which drives the op_code bus and selects the active digit modulus on the mod_select bus. Processing of a selected modulus affects the remaining digit moduli through crossbar buses internal to each register processor.

The divisor crossbar bus transmits the selected digit value during mixed-radix operations, while the dividend crossbar bus transmits the selected offset value during scaling and mixed-radix operations. In the present design, one recombination register base extends both the dividend and divisor. A future design could perform divisor decomposition and base extension once, store the resulting modulus sequence in a LUT, and reuse it for each iteration of (46).

Both crossbar buses also connect to a compare block, enabling comparison during simultaneous mixed-radix conversion. At the end of the divide process, the quotient and remainder are produced by the recombination/ALU unit.

## Initial FPGA Synthesis Results

The example RTL design provided as supplementary material was synthesized and implemented on a Cyclone IV E FPGA (part # EP4CE115F29C8) using Quartus version 18.1.0. This FPGA is featured on the widely used DE2-115 development board, a platform popular in academic settings. To isolate the performance of the RNS divide unit, the synthesis test was conducted using a virtual pin configuration. The FPGA resource utilization of the design is detailed in Figure 15.

| | |
|---|---|
| Fitter Status | Successful - Mon Dec 30 20:49:35 2024 |
| Quartus Prime Version | 18.1.0 Build 625 09/12/2018 SJ Standard Edition |
| Revision Name | MX9_divide_test4 |
| Top-level Entity Name | mx9_int_divide |
| Family | Cyclone IV E |
| Device | EP4CE115F29C8 |
| Timing Models | Final |
| Total logic elements | 27,106 / 114,480 ( 24 % ) |
| Total registers | 1612 |
| Total pins | 0 / 529 ( 0 % ) |
| Total virtual pins | 653 |
| Total memory bits | 100,608 / 3,981,312 ( 3 % ) |
| Embedded Multiplier 9-bit elements | 54 / 532 ( 10 % ) |
| Total PLLs | 0 / 4 ( 0 % ) |

*Figure 15 - Synthesis results using Cyclone IV FPGA*

Note the design consumes a relatively high percentage (24%) of logic elements. This is due in part to the synthesis of MOD and divide functions used in the DPU blocks to perform modular arithmetic as well as to detect or force divisibility by any power of base modulus for the divisor and dividend DPU respectively.

The number of 9-bit multipliers required is 10% to support parallel RNS multiplication for the two 18-digit RNS register processor blocks since inverse multiplication is executed for all digits simultaneously. Total LUT memory to store multiplicative inverses is only 3% of the available resources.

Synthesis for the design was relatively quick and dirty. No optimization of MOD or divide functions was performed, and only basic mod (%) and divide (/) function primitives were used for RTL synthesis. For this reason and many others, it is expected the division algorithm resource usage can be significantly improved.

## Initial FPGA Clock Speed

The clock speed of the initial RNS integer divide for the synthesis results above is 16.8 MHz. For implementation in the MOD-9 RNS ALU by Maitrix, LLC, which operates at approximately 40 MHz using the Cyclone IV FPGA, this speed is somewhat slower than hoped for. However, it is believed the RNS integer divide algorithm speed can be improved because of the many opportunities for optimizations. A full simulation display is shown in Figure 16.

# Simulation Snapshot

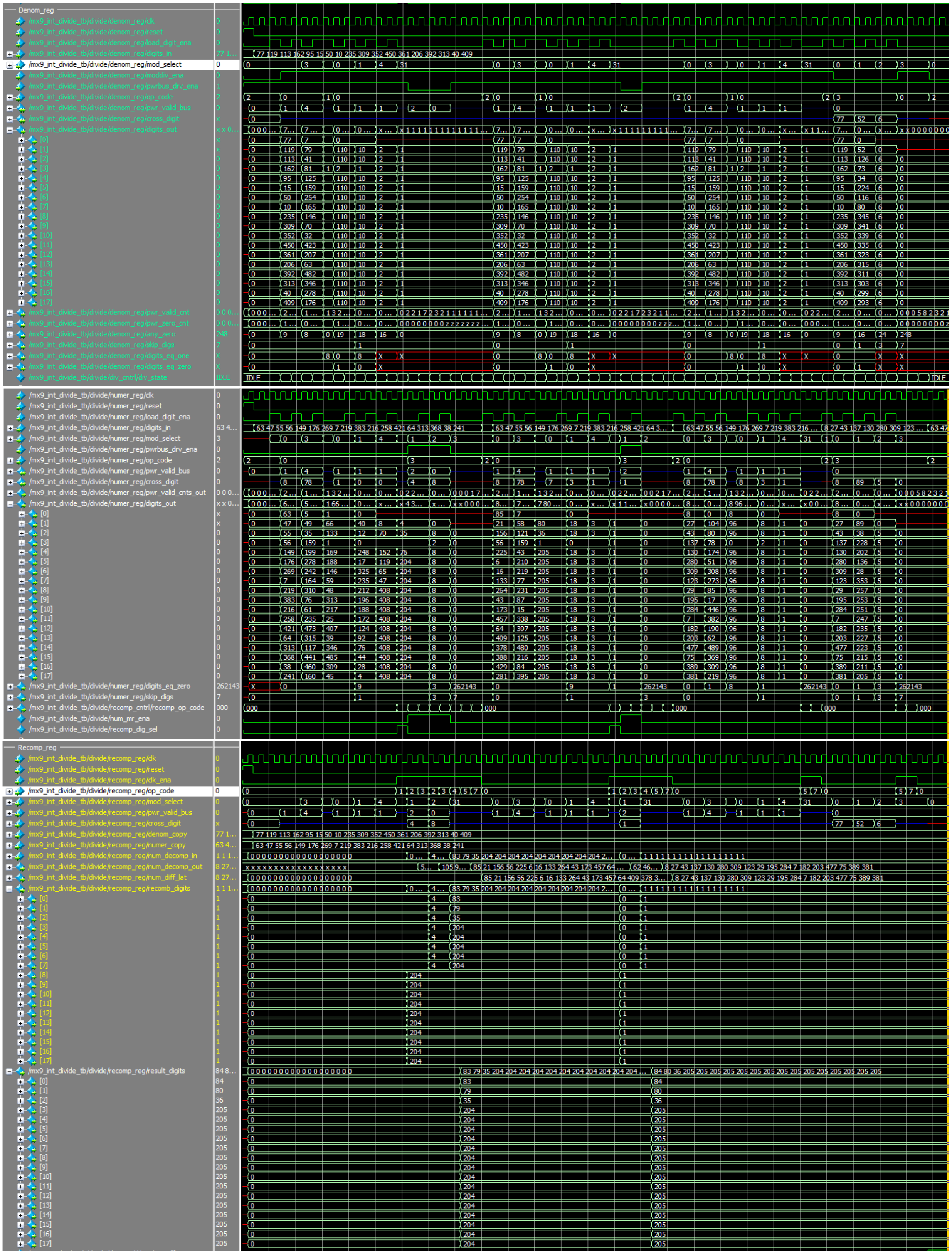

*Figure 16 - Simulation waveform including DENOM, NUMER, & RECOMP registers*

# Future Hardware and Algorithmic Enhancements

This section discusses future hardware and algorithmic enhancements. Unless explicitly stated otherwise, the items described here are proposals for future work and are not part of the implemented RTL design or the synthesis results reported in this paper.

## Parallel Apparatus Improvements

There is often a tradeoff between hardware resources and speed of operation. In the division algorithm presented, there are several key areas where additional hardware resources can be operated in parallel and will speed execution of division.

### Parallel Comparison

In the division algorithm presented, only a single comparison is required to resolve the final case statement of (47). In the code, the division algorithm re-uses the tri-state buses to perform comparison to save logic resources. One proposal is to execute a comparison with separate hardware upon each iteration of (45). When the condition that the NUMER register goes to zero during the DIV_DENOM state, the quotient result can be adjusted quickly saving some clock cycles.

Other related improvements include finding ways to eliminate the final comparison all together. For example, by integrating a comparison into the divisor decomposition algorithm. Investigations are required to determine if this is feasible.

### Recombination

Recombination of RNS digits is performed during base extension. In the code, when both the divisor and dividend register require base extension, this is performed in sequence. It is possible to include a second recombination unit so that this is performed in parallel.

## Algorithm Implementation Enhancements

There are many enhancements to the architecture and RTL design that will improve speed and efficiency of the RNS division algorithm. The following are some ideas.

### Divisor Base Extension

The divisor decomposition algorithm creates the same sequence of values for each iteration of (45) to represent the product $\hat{Y}$. Therefore, there is no reason to perform the divisor decomposition and any possible divisor base extension operation more than once. Upon the first full divisor decomposition, each scaling factor (base modulus and power) is stored in a small memory. For each iteration of (45), the factors of $\hat{Y}$ are read from memory to decompose the numerator. While this does not eliminate the need for clock cycles to decompose the numerator, it can eliminate the cycles for base extension of the divisor after the first divisor base extensions have been performed.

### Multi-Factor Scaling Enhancements

Because the set of factors comprising $\hat{Y}$ is known after the first divisor decomposition, it is possible to perform grouping of the factors comprising a single scaling operation. For example, combinations of base modulus, like m=2 and m=5, can be combined to scale by the value of 10 in a single scaling cycle. The complication arises in finding the minimum value to subtract from the numerator to perform the single scaling operation. However, the subtraction values can be computed, so it is up to the RTL designer to solve such problems, so they perform faster than the sequential scaling used herein. Note not all combinations need to be supported, but alternatively some of the most common combinations of factors can be supported. Note this optimization requires larger LUTs for storing more multiplicative inverses, or they can be computed on the fly and stored in temporary LUTs during the first divisor decomposition.

## Enhanced Design for MOD Functions

A basic enhancement technique is the optimization of the MOD function look-up tables. In the code, these MOD functions were implemented using the percentage (%) operator in Verilog. The resulting logic is synthesized using standard logic cells. Alternatively, MOD functions using customized LUTs with BRAM or other dedicated memory can yield faster performance thereby increasing clock speed.

Other techniques for calculating fast MOD function (with respect to a constant) include a method described in [11]. In this method, instead of representing each digit using binary, power-based digits use a binary coded fixed radix (BCFR) representation where the radix is equal to the base modulus. Detection of the divisibility of the denominator by one or more powers of the base modulus is made very quickly by detecting zeros in the least significant digit positions. (This is the origin of the term “zeros” to indicate divisibility of the divisor by its base modulus). This format also makes it easier to calculate the offset to be subtracted from the numerator during the divisor decomposition. Note this method requires conversion of BCFR representation to binary and binary to BCFR format to communicate the digit value to other digit processing units.

It should be noted the base modulus m=2 does not require LUTs since simple inspection of the binary bits themselves will indicate divisibility by one or more powers of two. The code does not partition this modulus out for special treatment currently. Still other methods to efficiently calculate the MOD function of a digit value with respect to a constant may be found in other prior works.

## Pipelining Hardware

Clock speed can be enhanced in many ways. Of particular interest is reducing the fanout of the tri-state buses. One idea is to isolate each tri-state bus using a single pipeline register between power-based moduli and non-power-based moduli. One complication arises from the increase in latency for detection of digit states, such as zero detection, etc. It may be possible to overcome or justify these latency increases by increased clock speed, or other techniques.

Another area for pipelining is the MOD functions. It should be noted the divisibility of any digit is not affected by inverse multiplication from the division of another digit. In other words, dividing by one factor does not change the divisor’s divisibility by another factor. By latching the state of the divisor, extra time for MOD function propagation can help speed the detection of divisibility, or detection of “zeros”. However, this is not the case for the dividend, which must undergo a subtraction by some value; this value is affected by each division. However, it is beneficial to speed detection of the divisibility of the divisor, as this propagates to control logic affecting the dividend.

## Delaying Base Extension

There are several strategies for delaying base extension. Since base extension is a multi-cycle operation, minimizing the number of times base extension is executed will reduce the number of cycles to complete the division. Some strategies are presented:

In one strategy, after the first divisor decomposition is performed, it is then known when base extension is needed for base modulus two for each scaling iteration. This knowledge can reduce the number of times the NUMER register is base extended in subsequent scaling iterations by intelligently selecting the most efficient points to execute a base extension operation. For some sets of data, optimizing when base extension is applied can significantly impact division efficiency.

In another strategy, the base extension occurring directly before the NUMER register is added to the ACCUM register can be eliminated, provided it is known that the non-standard RNS representation of the ACCUM register has the appropriate range to handle the summation. This can be handled by using a redundant digit. While redundant digits are not required for the algorithm presented, and in fact, have been avoided, it is easy to generate a redundant digit during the first base extension cycle that may be required during repeated scaling. In this case, summations of the ACCUM register will be in a non-normalized state. It should be noted that since the pattern of modulus reduction is always the same, the non-standard ACCUM format will be the same for each term of (47). However, a final base extension is still required in most instances, as the final quotient is assumed to be returned in a fully normalized RNS format.

## Theoretical Algorithm Enhancements

There are many opportunities to research the division algorithm and theory to enhance efficiency of RNS division. This may include new operations in RNS processing, such as employing CRT, or may include enhancements to the algorithmic techniques presented herein. The following are several ideas.

### Enhancing Divisor Decomposition

There are opportunities to increase the efficiency of RNS division by modifying the divisor decomposition algorithm. One attractive area of research is finding a better approximation of $\hat{Y}$ to $Y$. The closer these two values, the smaller number of iterations of (45) are required. In fact, it may be possible to rival binary division in many cases with such an enhancement. For example, there is no reason why the algorithm should always increment the divisor when there is no divisibility by a valid base modulus as expressed by (38). It is possible to decrement the divisor when additional rules are followed. This method was first explained in [11].

If the divisor is always decremented, then $\hat{Y} \leq Y$, so the iteration value $YZ_i$ now over-shoots the target quotient $X_{i-1}$ in (45). This means the accumulated quotient portion will be subtracted on the next iteration. When decrementing the divisor for all cases, the accumulated quotient portion will be alternately added then subtracted for each iteration in (45).

One opportunity to enhance division is to detect if a decrement of the divisor versus an increment of the divisor will result in a larger scaling. In practice, to completely divide out the largest number represented by the range of the word size, it is important to initially increment the divisor, as this guarantees there is no overflow of the initial iterated term, thereby eliminating the need to compare at this step. However, it is then possible to decrement the divisor if more divisibility is detected. Note the iteration step will need to be adjusted to add and/or subtract the portion of the quotient calculated depending on when a decrement was made. It is possible that a comparison is required for each iteration. The advantage of alternating the divisor increment and decrement is an area of potential study and may be desirable for the division of very large numbers.

# Conclusion

This paper has presented an improved method for direct arbitrary integer division in Residue Number Systems and a corresponding hardware architecture suitable for practical implementation. Building on the type-II division approach of Szabo and Tanaka, the work reformulates the division process around a power-based RNS, multi-factor scaling, mixed-radix conversion, and a new divisor decomposition method. Together, these elements provide a more structured and hardware-oriented path for performing quotient and remainder operations directly in RNS without conversion to binary.

Several contributions were made.  First, the use of power-based moduli increases representational efficiency and expands the set of low-value factors available to the division process. Second, the paper formalized scaling by one or more powers of a base modulus, including the associated treatment of digit invalidation and recovery. Third, mixed-radix conversion was adapted not only for comparison, but also for simultaneous recovery of multiple invalid digits during base extension. Fourth, the divisor decomposition method introduced here provides a practical means of generating an approximate divisor composed only of supported base-modulus factors, thereby enabling the iterative recurrence to proceed efficiently in hardware.

The paper also showed that these mathematical ideas map naturally into a digit-parallel hardware organization. The proposed digit processing units, recombination units, control structure, and supporting lookup-table methods demonstrate that direct RNS division is not only a theoretical possibility, but a realizable arithmetic function in FPGA-based hardware. Initial synthesis results confirm feasibility, while also indicating that substantial optimization opportunities remain in modular reduction, divisibility detection, base extension scheduling, and decomposition reuse.

More broadly, this work helps address one of the long-standing barriers to more complete general-purpose computation in RNS. While the operations of addition, subtraction, and multiplication are naturally efficient in RNS, direct division has remained comparatively immature. By developing both the algorithmic and hardware foundations for arbitrary integer division, this paper advances the practicality of RNS as a more capable computational number system.

Future work should focus on improving divisor approximation quality, reducing iteration count, refining base-extension strategy, optimizing hardware resource usage and clock speed, and exploring alternative decomposition and comparison methods. These directions may lead to substantially faster and more compact implementations and may further broaden the role of RNS in high-performance and specialized computing systems.

# Bibliography

[1] P. V. Ananda Mohan, *Residue Number Systems: Theory and Applications*, 1st ed., Cham, Switzerland: Birkhäuser, 2016.

[2] A. Nannarelli and M. Re, *Residue Number Systems: A Survey*, Technical University of Denmark, DTU Informatics, Technical Report No. 2008-04, 2008.

[3] M. A. Hitz and E. Kaltofen, "Integer Division in Residue Number Systems," *IEEE Transactions on Computers*, vol. 44, no. 8, pp. 983–989, 1995.

[4] N. S. Szabo and R. I. Tanaka, *Residue Arithmetic and Its Applications to Computer Technology*, New York, NY, USA: McGraw-Hill, 1967.

[5] A. Andraos, "Fixed Point Unsigned Fractional Representation in Residue Number System," in *Proceedings of the IEEE International Symposium on Circuits and Systems (ISCAS)*, 1997.

[6] E. B. Olsen, "RNS Hardware Matrix Multiplier for High Precision Neural Network Acceleration: 'RNS TPU'," in *Proceedings of the IEEE International Symposium on Circuits and Systems (ISCAS)*, 2018.

[7] E. B. Olsen, "Introduction of the Residue Number Arithmetic Logic Unit with Brief Computational Complexity Analysis," *arXiv preprint* arXiv:1512.00911, 2015.

[8] K. H. Rosen, *Elementary Number Theory and Its Applications*, 6th ed., Boston, MA, USA: Pearson, 2010.

[9] A. Kobin, *Elementary Number Theory*, course notes, 2023. [Online]. Available: https://static1.squarespace.com/static/5aff705c5ffd207cc87a512d/t/65cd0b38e6c4a358d42542c7/1707936569101/Elementary+Number+Theory.pdf

[10] R. P. Brent and P. Zimmermann, *Modern Computer Arithmetic*, Cambridge, UK: Cambridge University Press, 2010.

[11] E. B. Olsen, "Residue Number Arithmetic Logic Unit," U.S. Patent 9,081,608 B2, Jul. 14, 2015.

[12] B. Parhami, "Application of Symmetric Redundant Residues for Fast and Reliable Arithmetic," in *Proceedings of SPIE: Advanced Signal Processing Algorithms, Architectures, and Implementations XII*, vol. 4791, 2002.

# Appendix A: Table of Multiplicative Inverses

*Table 6 - Table of multiplicative* inverses *for power-based moduli used in examples.*

| | | WITH RESPECT TO MODULUS: | | | | | | | | | | | | | | | | | | | | | | | | | |
|---|---|---|---|---|---|---|---|---|---|---|---|---|---|---|---|---|---|---|---|---|---|---|---|---|---|---|---|
| | | **11** | **121** | **5** | **25** | **125** | **13** | **169** | **3** | **9** | **27** | **81** | **243** | **2** | **4** | **8** | **16** | **32** | **64** | **128** | **256** | **17** | **289** | **7** | **49** | **343** | **19** | **361** |
| INVERSE OF MODULUS | **11** | UND | UND | 1 | 16 | 91 | 6 | 123 | 2 | 5 | 5 | 59 | 221 | 1 | 3 | 3 | 3 | 3 | 35 | 35 | 163 | 14 | 184 | 2 | 9 | 156 | 7 | 197 |
| | **121** | UND | UND | 1 | 6 | 31 | 10 | 88 | 1 | 7 | 25 | 79 | 241 | 1 | 1 | 1 | 9 | 9 | 9 | 73 | 201 | 9 | 43 | 4 | 32 | 326 | 11 | 182 |
| | **5** | 9 | 97 | UND | UND | UND | 8 | 34 | 2 | 2 | 11 | 65 | 146 | 1 | 1 | 5 | 13 | 13 | 13 | 77 | 205 | 7 | 58 | 3 | 10 | 206 | 4 | 289 |
| | **25** | 4 | 92 | UND | UND | UND | 12 | 142 | 1 | 4 | 13 | 13 | 175 | 1 | 1 | 1 | 9 | 9 | 41 | 41 | 41 | 15 | 185 | 2 | 2 | 247 | 16 | 130 |
| | **125** | 3 | 91 | UND | UND | UND | 5 | 96 | 2 | 8 | 8 | 35 | 35 | 1 | 1 | 5 | 5 | 21 | 21 | 85 | 213 | 3 | 37 | 6 | 20 | 118 | 7 | 26 |
| | **13** | 6 | 28 | 2 | 2 | 77 | UND | UND | 1 | 7 | 25 | 25 | 187 | 1 | 1 | 5 | 5 | 5 | 5 | 69 | 197 | 4 | 89 | 6 | 34 | 132 | 3 | 250 |
| | **169** | 3 | 58 | 4 | 4 | 54 | UND | UND | 1 | 4 | 4 | 58 | 220 | 1 | 1 | 1 | 9 | 25 | 25 | 25 | 153 | 16 | 118 | 1 | 29 | 274 | 9 | 47 |
| | **3** | 4 | 81 | 2 | 17 | 42 | 9 | 113 | UND | UND | UND | UND | UND | 1 | 3 | 3 | 11 | 11 | 43 | 43 | 171 | 6 | 193 | 5 | 33 | 229 | 13 | 241 |
| | **9** | 5 | 27 | 4 | 14 | 14 | 3 | 94 | UND | UND | UND | UND | UND | 1 | 1 | 1 | 9 | 25 | 57 | 57 | 57 | 2 | 257 | 4 | 11 | 305 | 17 | 321 |
| | **27** | 9 | 9 | 3 | 13 | 88 | 1 | 144 | UND | UND | UND | UND | UND | 1 | 3 | 3 | 3 | 19 | 19 | 19 | 19 | 12 | 182 | 6 | 20 | 216 | 12 | 107 |
| | **81** | 3 | 3 | 1 | 21 | 71 | 9 | 48 | UND | UND | UND | UND | UND | 1 | 1 | 1 | 1 | 17 | 49 | 49 | 177 | 4 | 157 | 2 | 23 | 72 | 4 | 156 |
| | **243** | 1 | 1 | 2 | 7 | 107 | 3 | 16 | UND | UND | UND | UND | UND | 1 | 3 | 3 | 11 | 27 | 59 | 59 | 59 | 7 | 245 | 3 | 24 | 24 | 14 | 52 |
| | **2** | 6 | 61 | 3 | 13 | 63 | 7 | 85 | 2 | 5 | 14 | 41 | 122 | UND | UND | UND | UND | UND | UND | UND | UND | 9 | 145 | 4 | 25 | 172 | 10 | 181 |
| | **4** | 3 | 91 | 4 | 19 | 94 | 10 | 127 | 1 | 7 | 7 | 61 | 61 | UND | UND | UND | UND | UND | UND | UND | UND | 13 | 217 | 2 | 37 | 86 | 5 | 271 |
| | **8** | 7 | 106 | 2 | 22 | 47 | 5 | 148 | 2 | 8 | 17 | 71 | 152 | UND | UND | UND | UND | UND | UND | UND | UND | 15 | 253 | 1 | 43 | 43 | 12 | 316 |
| | **16** | 9 | 53 | 1 | 11 | 86 | 9 | 74 | 1 | 4 | 22 | 76 | 76 | UND | UND | UND | UND | UND | UND | UND | UND | 16 | 271 | 4 | 46 | 193 | 6 | 158 |
| | **32** | 10 | 87 | 3 | 18 | 43 | 11 | 37 | 2 | 2 | 11 | 38 | 38 | UND | UND | UND | UND | UND | UND | UND | UND | 8 | 280 | 2 | 23 | 268 | 3 | 79 |
| | **64** | 5 | 104 | 4 | 9 | 84 | 12 | 103 | 1 | 1 | 19 | 19 | 19 | UND | UND | UND | UND | UND | UND | UND | UND | 4 | 140 | 1 | 36 | 134 | 11 | 220 |
| | **128** | 8 | 52 | 2 | 17 | 42 | 6 | 136 | 2 | 5 | 23 | 50 | 131 | UND | UND | UND | UND | UND | UND | UND | UND | 2 | 70 | 4 | 18 | 67 | 15 | 110 |
| | **256** | 4 | 26 | 1 | 21 | 21 | 3 | 68 | 1 | 7 | 25 | 25 | 187 | UND | UND | UND | UND | UND | UND | UND | UND | 1 | 35 | 2 | 9 | 205 | 17 | 55 |
| | **17** | 2 | 57 | 3 | 3 | 103 | 10 | 10 | 2 | 8 | 8 | 62 | 143 | 1 | 1 | 1 | 1 | 17 | 49 | 113 | 241 | UND | UND | 5 | 26 | 222 | 9 | 85 |
| | **289** | 4 | 103 | 4 | 9 | 109 | 9 | 100 | 1 | 1 | 10 | 37 | 37 | 1 | 1 | 1 | 1 | 1 | 33 | 97 | 225 | UND | UND | 4 | 39 | 235 | 5 | 5 |
| | **7** | 8 | 52 | 3 | 18 | 18 | 2 | 145 | 1 | 4 | 4 | 58 | 139 | 1 | 3 | 7 | 7 | 23 | 55 | 55 | 183 | 5 | 124 | UND | UND | UND | 11 | 258 |
| | **49** | 9 | 42 | 4 | 24 | 74 | 4 | 69 | 1 | 7 | 16 | 43 | 124 | 1 | 1 | 1 | 1 | 17 | 17 | 81 | 209 | 8 | 59 | UND | UND | UND | 7 | 140 |
| | **343** | 6 | 6 | 2 | 7 | 82 | 8 | 34 | 1 | 1 | 10 | 64 | 226 | 1 | 3 | 7 | 7 | 7 | 39 | 103 | 103 | 6 | 91 | UND | UND | UND | 1 | 20 |
| | **19** | 7 | 51 | 4 | 4 | 79 | 11 | 89 | 1 | 1 | 10 | 64 | 64 | 1 | 3 | 3 | 11 | 27 | 27 | 27 | 27 | 9 | 213 | 3 | 31 | 325 | UND | UND |
| | **361** | 5 | 60 | 1 | 16 | 116 | 4 | 147 | 1 | 1 | 19 | 46 | 208 | 1 | 1 | 1 | 9 | 25 | 25 | 89 | 217 | 13 | 285 | 2 | 30 | 324 | UND | UND |

## Appendix B: RTL and C++ Code Materials

A supplemental RTL repository containing the synthesizable RTL code for arbitrary division on 150-bit RNS numbers can be accessed here https://github.com/MaitrixLLC/RNS-Integer-Divide-Supplement

A software repository containing the RNS APAL C++ source code can be found here: https://github.com/MaitrixLLC/RNS-APAL